\documentclass[oldversion]{aa}  
\usepackage{graphicx}
\usepackage{txfonts}

\newcommand{\diff}{\mbox{d}}
\newcommand{\beq}{\begin{equation}}
\newcommand{\eeq}{\end{equation}}
\newcommand{\beqa}{\begin{eqnarray}}
\newcommand{\eeqa}{\end{eqnarray}}
\newcommand{\benu}{\begin{enumerate}}
\newcommand{\eenu}{\end{enumerate}}
\newcommand{\bite}{\begin{itemize}}
\newcommand{\eite}{\end{itemize}}
\newcommand{\bdes}{\begin{description}}
\newcommand{\edes}{\end{description}}

\newcommand{\comment}[1]{}
\begin{document}
\title{{Scaled solar tracks and isochrones in a large region of 
the $Z$--$Y$ plane }
\subtitle{ II. From  $ 2.5$ to $ 20 M_{\odot}$ stars }}

\author{G. Bertelli\inst{1} \and E. Nasi\inst{1} \and L. Girardi\inst{1} \and P. Marigo\inst{2}
} 

\offprints{G. Bertelli, \email{gianpaolo.bertelli@oapd.inaf.it}}

\institute {INAF - Padova Astronomical Observatory, Vicolo 
dell'Osservatorio 5, 35122 Padova, Italy 
\and Astronomy Department, Padova University,
 Vicolo dell'Osservatorio 3, 35122 Padova, Italy}
    
\date{Received / accepted }

 
\abstract
{We extend our theoretical computations for low-mass stars to 
intermediate-mass and massive stars, for which few databases exist in the 
literature. Evolutionary tracks and isochrones are computed
for initial masses $2.50 - 20 M_{\odot}$ for a
grid of 37 chemical compositions with  metal content $Z$ between
0.0001 and 0.070 and helium content $Y$ between 0.23 and 0.40 to enable users
to obtain isochrones for ages as young as about $10^7$ years  and to simulate
stellar populations with different helium-to-metal enrichment laws. 
The Padova stellar evolution code is identical to that used in the first
paper of this series. Synthetic TP-AGB 
models allow stellar tracks and isochrones to be extended until the end of 
the thermal pulses along the AGB. We provide software tools for the 
bidimensional interpolation (in $Y$ and $Z$) of the isochrones.

We present tracks
for scaled-solar abundances and the corresponding 
isochrones from very old ages down to about $10^7$ years. This lower limit 
depends on chemical composition.
The extension of the blue loops and the instability strip of Cepheid stars are
compared and the Cepheid mass-discrepancy is discussed. The location of red 
supergiants in the H-R diagram is in good agreement with the evolutionary 
tracks for masses from $10$ to $20 M_{\odot}$. 
Tracks and isochrones are available in tabular form for the
adopted grid of chemical compositions in the extended plane $Z-Y$ in three 
photometric systems. An interactive web interface allows users to obtain
isochrones of any chemical composition inside the provided $Z-Y$ range 
and also to simulate stellar populations with different $Y(Z)$ helium-to-metal
enrichment laws.  }


  \maketitle
%

\section{Introduction}

In general, the helium content of stellar populations is poorly known. For
super-solar metallicities, the helium content is unconstrained, and has usually
been derived from simple reasoning, based on naive chemical evolution
models of galaxies, rather than observational grounds.

The most popular assumption in evolutionary models of stars is that $Y$ 
increases linearly with $Z$.
This assumption is inaccurate for old globular clusters, some of which contain 
multiple stellar populations with evidence of significant helium enrichment
(Piotto et al. 2007; Piotto 2009). Even the solar initial content of helium
has been debated, since the values derived from helioseismology depend on the
assumed efficiency of helium diffusion (Guzik et al. 2005, 2006; Basu \& Antia
2008), whereas on 
the other hand helioseismology has been challenged by the debate about the 
oxygen abundance (Montalban et al. 2004; Scott et al. 2009). For these reasons 
it is likely that the helium content of stellar 
evolutionary models will in the future be revised many times. An alternative 
approach, followed in this paper, is that of computing stellar models for a 
significant range of the $Z$ - $Y$ plane, so that interpolations in the grid 
allow us to consider many possible $Y(Z)$ relations.

A large amount of theoretical investigation has been devoted to the evolution
of low-mass stars populating old galactic globular clusters, since rich stellar
clusters provide us with the opportunity to test the results of stellar 
evolution theories. The increasing amount of data for young populous clusters 
in both of the
Magellanic Clouds encourages us to pay special attention to the evolutionary 
behaviour
of  more massive stars presently evolving in those clusters. However,
not so much theoretical computation has been addressed to the implementation 
of existing stellar model databases for more massive stars, as done for low 
masses.
 
The Padova and Geneva groups originally compiled grids of stellar
models with overshooting up to $120 M_{\odot}$ for several chemical 
compositions (Bressan et al.
1993; Fagotto et al. 1994a,b, and relative isochrones in Bertelli et al. 1994;
Schaller et al. 1992; Schaerer et al. 1993a,b; Charbonnel et al. 1993; Meynet 
et al. 1994).
Girardi et al. (2000) computed a new set of low- and intermediate-mass stellar
models with updated opacities and equation of state. 
Bono et al. (2000) presented intermediate-mass standard models with different
helium and metal content ($ 3 \le M \le 15 M_{\odot}$), and Pietrinferni et al.
(2004, 2006) database makes available stellar models and isochrones
for scaled-solar  and $\alpha$-enhanced metal distributions in the mass range 
between $0.5$ and $10 M_{\odot}$ with and without overshooting.

Following the first release of the new stellar evolution models of the Padova 
database for low-mass stars in a large region of the $Z$-$Y$ plane by Bertelli 
et al. (2008) (hereinafter Paper I), in this paper we present models for stars 
from $2.5$ to $20 M_{\odot}$ with the same range of chemical compositions as 
in Paper I. Among the astrophysical problems of interest in this
range of mass, we discuss whether observations of Cepheids in the Magellanic 
Clouds (MCs) and the Galaxy can be reproduced by
these new models. Massive red supergiants in the Galaxy and in MCs with newly
derived physical parameters by Levesque et al. (2005, 2006) are compared
with our models from $10$ to $20 M_{\odot}$, evaluating the adequacy of
stellar models in this mass range.

A neglected process in our massive stellar models is stellar rotation. 
Significant advances have been made in this field  (e.g., Heger \& 
Langer 2000; Meynet \& Maeder 2000), which demonstrated
that the evolutionary paths of rapidly rotating massive stars can
differ from those of non-rotating stars. A book by Maeder (2009) on the 
physics, formation, and evolution of rotating stars presents a very detailed 
description of the effects of rotation on stellar evolution.
Rapid rotation can trigger strong mixing inside massive stars, extend the 
core hydrogen-burning lifetime, significantly increase the luminosity, and
change the chemical composition at the stellar surface with time.
Regardless of the success of rotating stellar models in explaining a variety of
stellar phenomena, rotational mixing is still a matter of debate (de Mink et
al., 2009). 

Our choice of neglecting stellar rotation in this set of models has two main
motivations, namely: 

1) not all stars begin their life with the same rotational velocity; 

2) stars have rotational axes that are not all oriented in the same direction. 

These two stochastic effects introduce some dispersion in the
distribution of the stars in the HR disgram, in contradiction with the 
deterministic character of stellar isochrones.  The ``non-rotating
isochrones'' represent the starting point for any further study.

The present paper is organized as follows.
In Sect. 2, we summarize the main input physics and in Sect. 3,  TP-AGB 
models are described. In  Sect. 4, stellar tracks are presented with a
description of tables indicating changes in surface chemical
composition after the first and the second dredge-up.
Isochrones and relative tables are described in Sect. 5, in addition to the
interpolation scheme allowing to obtain isochrones and to simulate stellar
populations in a large region of the $Z-Y$ plane.
Cepheids and the problem of the mass discrepancy are discussed in Sect. 6.
Section 7 deals with massive red supergiants  in the Galaxy and the 
Magellanic Clouds, and Sect. 8 presents the concluding remarks.

\section {Input physics and coverage of the $Z-Y$ plane}

The stellar evolution code adopted in this work is the same
as in Bertelli et al. (2008, hereinafter Paper I ). 
The  masses described in this paper are between $2.5$ and $20 
M_{\odot}$,  to ensure that we have a complete stellar model database from 
$0.15$ to $20 M_{\odot}$.
In the following, we  summarize the input physics described in Paper I.

\subsection{Initial masses and chemical compositions}
\label{sec_chemic}

\begin{table}[!ht]
\caption{Combinations of $Z$ and $Y$ of the computed tracks}
\label{tab_comp}
\smallskip
\begin{center}
{\small
\begin{tabular}{cccccc}
\hline
\noalign{\smallskip}
Z & Y1 & Y2 & Y3 & Y4 & Y5 \\
\noalign{\smallskip}
\hline
\noalign{\smallskip}
0.0001 & 0.23 & 0.26 & 0.30 &      & 0.40      \\
0.0004 & 0.23 & 0.26 & 0.30 &      & 0.40      \\
0.001  & 0.23 & 0.26 & 0.30 &      & 0.40      \\
0.002  & 0.23 & 0.26 & 0.30 &      & 0.40      \\
0.004  & 0.23 & 0.26 & 0.30 &      & 0.40      \\
0.008  & 0.23 & 0.26 & 0.30 & 0.34 & 0.40      \\
0.017  & 0.23 & 0.26 & 0.30 & 0.34 & 0.40      \\
0.040  &      & 0.26 & 0.30 & 0.34 & 0.40      \\
0.070  &      &      & 0.30 & 0.34 & 0.40      \\
\noalign{\smallskip}
\hline
\end{tabular}
}
\end{center}
\end{table}
The first release of the new evolutionary models was presented
in  Paper I for masses from $0.15 $  to $2.5 M_{\odot}$ (Bertelli et al.
2008). 
The stellar models were computed 
for initial masses $2.5, 3, 3.5, 4, 4.5, 5, 6, 7, 8, 10, 12, 15$, and $20 
M_{\odot}$,
and, as in Paper I, from the ZAMS to the end of helium burning.
The initial chemical composition is in the range $0.0001
\le Z \le 0.070$ for the metal content, and for the helium content in
the range $0.23 \le Y \le 0.40$ as shown in Table 1. In addition, tracks were
computed for masses between $4.5$ and $20 M_{\odot}$ with $Z=0.030$  and 
$Y=0.26, 0.30, 0.34,$ and $0.40$ to derive more reliable interpolated 
isochrones for more massive stars.

\subsection{The choice of scaled-solar abundances }

For each value of $Z$, the fractions of different metals follow a
scaled solar distribution, as compiled by Grevesse \& Noels (1993), and
adopted in the OPAL opacity tables. The ratio of abundances of
different isotopes is assumed to be identical to Anders \& Grevesse (1989).

In this regard, we note that
both the solar composition, and what is referred to as the alpha-enhanced
distribution of metals have changed, by factors of up 
to 0.3 dex, in the past few years.
Moreover, there is also an increasing perception that neither scaled solar, nor
alpha-enhanced compositions are universal: they  
apply only to particular regions of spiral galaxies that are similar to our 
own. For instance, there are documented cases of slightly alpha-depleted 
composition in some areas of the 
LMC (e.g., Pomp\'eia et al. 2008). For nearby
dwarf galaxies, the modern observational picture is that their populations are 
in general slightly alpha-depleted for [Fe/H]$>-1.0$, becoming alpha-enhanced only at very low
metallicities (see Fig. 11 in Tolstoy et al. 2009, and references therein).  
On the high-metallicity end, although it is generally believed that the most 
massive ellipticals are metal-rich and alpha-enhanced, there is no guarantee 
that they follow the same distribution of alpha-elements as  nearby halo 
stars.

The above-mentioned observations of dwarf galaxies demonstrate that 
low-metallicity models with scaled solar compositions are useful and, in
many cases, appropriate, to the study of nearby galaxies. Since the 
observational view of  a ``typical alpha-enhanced composition'' continues to
change, we prefer to be conservative and begin the modelling from the most well
studied case, the scaled solar one, which is reasonably well-known, apart from
the solar oxygen abundance. We will explore a few possible cases of 
alpha-enhancement in future papers.This approach is consistent with our choice 
of scaled solar abundances and the rules 
introduced by Salaris et al. (1993) to approximate low-Z alpha-enhanced
isochrones using scaled solar ones (which can be applied to
the Milky Way halo, metal-poor globular clusters, and the metal-poorest 
stars in dwarf galaxies).    

As far as the use of  ``old solar abundances'' is concerned (Grevesse \& Noels,
 1993), we note the following:
\begin{itemize}
\item
Some more recent studies (e.g., Caffau et al. 2008, and references therein)
confirm  the high O abundances typical of older compilations of solar
abundances, rather than the low values claimed by 
Asplund et al.(2004). Solar models computed with low heavy-element 
abundances also strongly disagree with the constraints from helioseismology 
(see Basu \& Antia 2008, and references therein).

\item 
We computed a few sequences of stellar tracks using several 
 solar compositions. The difference 
between the models computed at the same mass, metallicity, and mixing length
parameter was found to be negligible  (as discussed in Sect. 4.3.1) .
\end{itemize}  
 
\subsection{Opacities}
\label{sec_opac}

The radiative opacities for scaled solar mixtures are assumed to be those of
the OPAL group (Iglesias \& Rogers 1996) for temperatures higher than $\log T =
4$, and the molecular opacities are those of Alexander \& Ferguson (1994) for
$\log T < 4.0$ as in Salasnich et al. (2000).
 For very high temperatures ($\log T \ge 8.7$), we use the
opacities of Weiss et al. (1990).

The conductive opacities of electron-degenerate matter are taken from Itoh
et al. (1983). For the
interpolation within the opacity tables grids in both papers, we
used the two-dimensional bi-rational cubic damped-spline algorithm
(see Schlattl \& Weiss 1998; Salasnich et al. 2000, and Weiss \&
Schlattl 2000). In general, evolutionary models of low-mass stars differ very 
little by changing the opacity interpolation scheme. The model predictions for
massive stars ($M \ge 5 
M_{\odot}$) are far more sensitive to the interpolation algorithm as
discussed in Salasnich et al. (2000)

\subsection{Equation of state}
\label{sec_eos}

The equation of state (EOS) for temperatures higher than $10^7$~K is
that of a fully-ionized gas, including electron degeneracy in the way
described by Kippenhahn et al.\ (1967). The effect of Coulomb
interactions between the gas particles at high densities is taken into
account as described in Girardi et al.\ (1996).
For temperatures lower than $10^7$~K, the detailed ``MHD'' EOS of
Mihalas et al.\ (1990, and references therein) is adopted. 
We note that the MHD EOS is critical only for stellar models of
mass lower than 0.7~ $M_{\odot}$ during their main sequence evolution.

\subsection{Reaction rates and neutrino losses}
\label{sec_rates}

The reaction rates are taken from the compilation of Caughlan \& Fowler
(1988, hereinafter CF88), apart from $^{17}{\rm O}({\rm
p},\alpha)^{14}{\rm N}$ and $^{17}{\rm O}({\rm p},\gamma)^{18}{\rm
F}$, for which we use the determinations by 
Landr\'e et al.\ (1990). The
uncertain $^{12}$C($\alpha,\gamma$)$^{16}$O rate was set to be 1.7 times
the values given by Caughlan \& Fowler (1988), as indicated by the
study of Weaver \& Woosley (1993) of the nucleosynthesis by massive
stars.  The electron screening factors for all reactions are those
from Graboske et al.\ (1973).
The abundances of the various elements are evaluated with the aid of a 
semi-implicit extrapolation scheme, as described in Marigo et al. (2001).
The energy losses by neutrinos are from Haft et al. (1994).

In Sect. 4.3.2 we discuss the effects on stellar models of
measurements of the $^{14}{\rm N}({\rm p},\gamma)^{15}{\rm O}$ reaction rate by
the LUNA experiment (Formicola et al., 2004).

\subsection{Convection}
\label{sec_conv}

The most important physical process for mixing in stars is convection, but at 
present convection remains a main  source of uncertainty in stellar models 
computations.
In our stellar models, convection is treated with the Schwarzschild 
criterion for stability.
Usually  energy transport in the outer convection zone of stars is described
according to the mixing-length theory (MLT) of B\"ohm-Vitense (1958).
We adopt the same value as in Paper I for the MLT parameter ($\alpha =1.68$),
calibrated by means of the solar model. 

\subsubsection{ Overshoot}
The extension of the convective regions in stellar models takes into account 
overshooting from the borders of both core and envelope convective zones 
(Bressan et al. 1981; Alongi et al. 1991; Girardi et al. 2000).  
In the following, we adopt the formulation by Bressan et al. (1981) in which
the boundary of the convective core is defined to be the layer where the 
velocity 
(rather than the acceleration) of convective elements vanishes. This
non-local treatment of  convection requires the use of a free parameter related
to  the mean free path $l$ of convective elements defined to be $l=\Lambda_c 
H_p$, ($H_p$ being the pressure scale-height).   The choice of this parameter
determines the extent of the overshooting  
{\em across} the border of the classical core (determined by the Schwarzschild
 criterion).
 Other authors define the extent
of the overshooting zone to be at the distance $d=\Lambda_c H_p$ {\em above} 
the border of the convective core (Schwarzschild criterion).   
The value 0.5 used in the Padova models is almost equivalent to the 0.2
value adopted by other groups (Meynet et al, 1994; Pietrinferni et al. 2004;
Yi et al. 2001). 
 The non-equivalence of the parameter
used to describe the extension of convective overshooting by different groups
 has been a recurrent source of misunderstanding in the literature.

We adopt the same prescription as in Paper I (Bertelli et al. 2008) and in
Girardi et al. (2000) for the parameter $\Lambda_c$  for H- and He-burning 
stages.
Overshooting at the lower boundary of convective envelopes is also considered. 
For $M>2.5 M_{\odot}$, a value of $\Lambda_e=0.7$ was assumed, as in
Bertelli et al.\ (1994) and Girardi et al. (2000).

\subsubsection{Hydrogen semiconvection} 
During the core H-burning phase of massive stars,
radiation pressure and electron scattering opacity produce a large 
convective core surrounded by an H-rich region, which is potentially unstable
to convection if the original gradient in chemical abundance remains, but 
stable if suitable mixing is allowed to occur. This region of the star
is assumed in theoretical models to undergo sufficient mixing until the
condition of neutrality is restored, to transport negligible energy flux. The
Schwarzschild condition produces smoother chemical profiles and, in some
cases,  a fully intermediate convective layer.
Similar instability also occurs during the early shell H-burning phase. Taking
into account mass loss from massive stars, the extension of semiconvection 
and/or intermediate full convection at the top of the H-burning shell is much 
smaller than in constant-mass stellar models.   
The effects of H-semiconvection on the evolution of massive stars were
summarized by Chiosi \& Maeder (1986).
 
\subsubsection{ Helium semiconvection} 
As He-burning proceeds in the convective core of
low-mass stars during the early stages of the horizontal branch (HB), 
the size of the convective core increases.
Once the central value of helium falls below $Y_c =0.7$, the 
temperature gradient reaches a local minimum, so that continued overshoot is
no longer able to restore the neutrality condition at the border of the core.
The core divides into an inner convective
core and an outer convective shell. As further helium is captured by the 
convective shell, the outer shell tends to become stable, leaving behind a 
region of varying composition in conditions of neutrality.
This zone is called semiconvective.
Starting from the centre and going outwards, the matter
in the radiatively stable region above the formal convective core is mixed 
layer by layer until neutrality is achieved. This picture holds 
during most of the central He-burning phase.
The extension of the semiconvective region
varies with stellar mass, being important in  low- and intermediate-mass
stars up to about $5 M_\odot$, and negligible in more massive stars.
We followed the scheme of Castellani et al. (1985), as described in Bressan
et al. (1993). 

\subsection {Mass loss }
\label{sec_rgbagb}

Mass loss has a crucial impact on the evolution of massive stars, affecting 
evolutionary tracks, lifetimes, and surface abundances.
The role of radiation pressure in driving mass loss from massive stars has
been firmly accepted by theoretical astrophysics since the pioneering studies 
by Lucy \& Solomon (1970), and Castor et al. (1975). A relevant review on the 
evolution of massive stars with mass loss was given by Chiosi \& Maeder (1986), while
Kudritzki \& Puls (2000) exhaustively describe winds from hot stars.
The metallicity dependence of mass loss rates is usually included using the 
formula: 

$\dot M(Z) = \dot M(Z_{\odot}) (Z/Z_{\odot})^{\alpha}$,

\noindent where the exponent $\alpha$ varies between  values of $0.5 - 0.6$ 
(Kudritzki \& Puls 2000; Kudritzki 2002), and $0.7 - 0.86$ (Vink et al.
 2001; Vink \& de Koter 2005) for O-type and WR stars. 
A comparison between mass loss prescriptions and observed mass loss 
rates, and the observational dependence of the rate on the metal content of
massive stars atmospheres was presented by Mokiem et al. (2007).
They found a metallicity dependence consistent with previous values and 
stated that those $\dot M(Z)$ relations hold for stars more luminous than 
$\sim 10^{5.2} L_{\odot}$.  

In this paper, we adopt the mass loss rates from stellar winds driven by 
radiation pressure according to 
de Jager et al.(1988) for all evolutionary stages starting from the main 
sequence and  from about $\sim 6 - 7 $
 to $20 M_{\odot}$ (the upper mass limit to this evolutionary track release). 
The rates incorporate the dependence on metallicity given by
Kudritzki et al. (1989)

  $\dot M_Z = (Z/Z_{\odot})^{0.5} \dot M_{Z_{\odot}}$.

In a future paper dealing with the evolution of massive stars, for the 
mass-loss rates   we will adopt relationships based on more recent 
observations and/or determinations.


\section{Synthetic TP-AGB models}

An important update of the database of evolutionary tracks is the extension
of stellar models and isochrones until the end of the thermal pulses  along
the asymptotic giant branch (TP-AGB) for all masses up to $\sim 6 M_{\odot}$.
The evolution from the first thermal pulse  to the complete ejection of
the stellar envelope is computed with the synthetic code described in Marigo
\& Girardi (2007, and references therein), to whom we refer for all details. 
It is important to mention that these TP-AGB tracks have been updated
in many aspects, describing for instance the transition to the C-star phase 
by means of the third dredge-up event, the prevention of this transition for 
the 
most massive tracks because of hot bottom burning, the proper effective 
temperatures 
of carbon stars, and suitable mass-loss rates of the M and C-type stars.
Although the models are not calibrated in the same way as Marigo \& Girardi 
(2007)
tracks (i.e., present models might not reproduce the luminosity functions and
lifetimes of carbon stars in the Magellanic Clouds), they represent a 
significant improvement to the simple way the TP-AGB tracks were included 
in many previous sets of isochrones.

\section{Stellar tracks}
\label{sec_tracks}

%
%


\begin{figure}
\resizebox{\hsize}{!}{\includegraphics{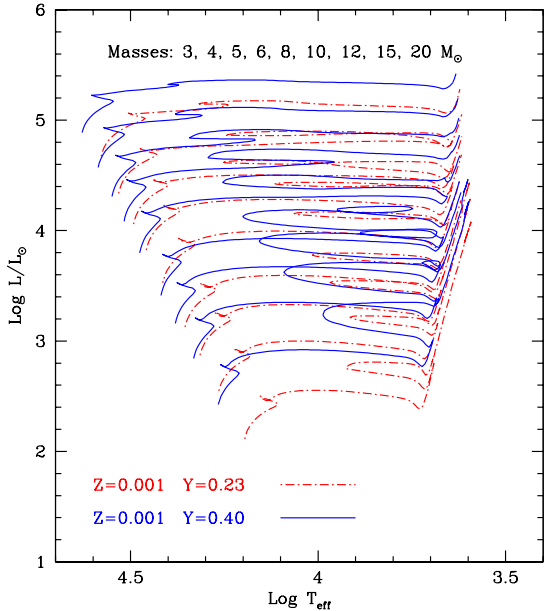}}

\caption{  
Evolutionary tracks in the HR diagram for the composition $Z=0.001,
Y=0.23$ (dot-dashed line) and $Z=0.001, Y=0.40$ (solid line). 
 }
\label{hrd_z001}
\end{figure} 

\begin{figure}
\resizebox{\hsize}{!}{\includegraphics{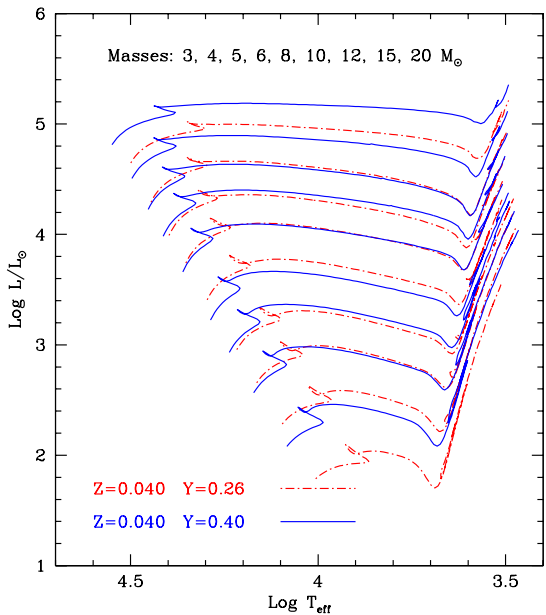}}
\caption{
Evolutionary tracks in the HR diagram for the composition $Z=0.040,
Y=0.26$ (dot-dashed line) and $Z=0.040, Y=0.40$ (solid line). 
 }

\label{hrd_z040}
\end{figure}

\subsection{Evolutionary stages and mass ranges}
\label{sec_massranges}
We consider that stellar models are on the zero age main-sequence (ZAMS) when 
chemically homogeneous stars begin their H-burning and the gravitational 
energy production is less than about 1 \% of the total energy.
Our models are evolved from the ZAMS, and the evolution is followed in detail 
through the entire H- and He-burning phases.
The tracks are stopped at the beginning of the TP-AGB phase (bTP-AGB)
in intermediate-mass stars, or at the carbon ignition in 
our more massive models. Stellar models and isochrones are extended until the 
end of the thermal pulses along the AGB with the synthetic TP-AGB models 
by Marigo \& Girardi (2007).

\subsection{Tracks in the HR diagram}
\label{sec_hrd}

We display 
only some examples of the sets of  computed evolutionary tracks 
(i.e., $Z=0.001$ for $Y=0.23$ and $Y=0.40$, 
and $Z=0.040$ for $Y=0.26$ and $Y=0.40$) for masses between $2.5$ and $20 
M_{\odot}$ from the ZAMS  to the beginning of either the TP-AGB, or  
 carbon burning.  
Figures 1 and 2 illustrate the range of 
temperatures and 
luminosities involved for the boundary values of helium content at a given 
metallicity in the theoretical HR diagram. These figures also illustrate the
dependences of the extension of blue loops on both metallicity and helium 
content, which are significant at low metallicity, and practically absent for 
higher metal content ($Z \ge 0.040$).
The extension of blue loops requires a detailed discussion since there are 
observations that are not easily reproduced by stellar models, such as 
Cepheids, 
and this topic is dealt with in Sect. 6. The location of red supergiants (RSG) 
on the H-R diagram did not always agree with the predictions of stellar 
evolutionary models, previousy  appearing to be too cool and too luminous to 
coincide with the position of the evolutionary tracks. New effective 
temperatures and luminosities for the RSG populations of galaxies by Levesque
et al. (2005, 2006) greatly improved the agreement between the RSGs and the 
evolutionary models, as is evident in Sect. 7 from the comparison with our new
moderate massive models.
The data tables for the present evolutionary tracks are available only
in electronic format.\footnote{ A website with the complete data-base 
(including additional data and future
extensions) will be maintained at {\bf http://stev.oapd.inaf.it/YZVAR}
with suitable information on electronic tables (the same as in Paper I).}

\subsection{Testing the effects of different input physics}

We computed a few sequences for different initial chemical compositions, 
opacities, and for nuclear reaction rates from the LUNA experiment, obtaining
very modest changes in the tracks, as discussed in the following.

\subsubsection{Initial chemical composition and opacities}

Updating both the chemical mixture (EOS) and the low-temperature opacities 
(determined with the AESOPUS code by Marigo \& Aringer, 2009), 
we computed a few sequences of stellar tracks using the distribution of 
element abundances of the solar chemical composition according to 
a) Grevesse \& Noels (1993), b) Grevesse \& Sauval (1998), and c) 
Grevesse, Asplund \& Sauval (2007). 
The differences between the models for the same $Z$ are negligible. 
To complete additional
comparison, we also used the opacity tables by Alexander \& Ferguson 
(1994), with reference to the Grevesse (1991) solar composition. We found 
that the differences between the models, computed at the same mass, 
metallicity and mixing-length parameter, are completely negligible. In 
particular, the corresponding giant branches in the H-R diagram practically 
overlap, the typical separation at fixed luminosity being of the order
of $0.002$ dex or less  throughout their extension.

\subsubsection{Nuclear reaction rates}
\begin{figure}
\resizebox{\hsize}{!}{\includegraphics{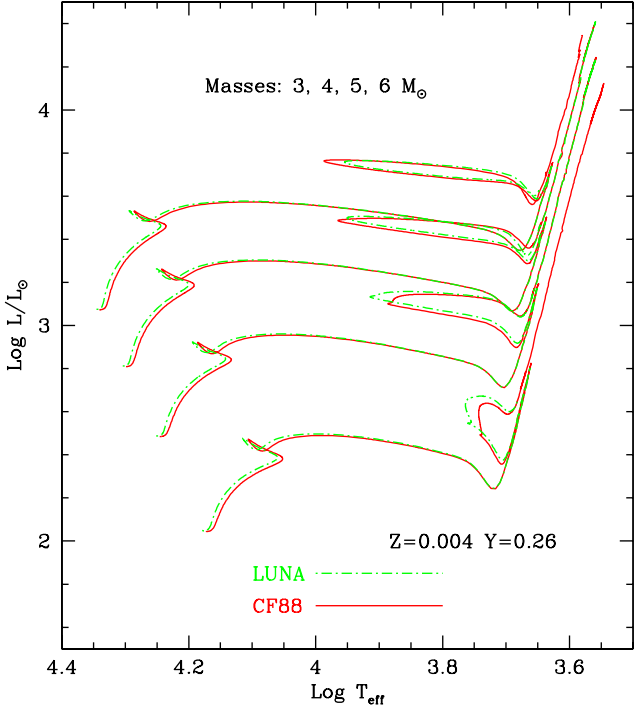}}

\caption{  
Evolutionary tracks in the HR diagram for the composition $Z=0.004,
Y=0.26$ computed with the new LUNA rate for the 
$^{14}{\rm N}({\rm p},\gamma)^{15}{\rm O}$ nuclear reaction (dot-dashed line) 
and with the rate according to Caughlan \& Fowler (1988)(solid line). 
}
\label{LUNA_z004}
\end{figure} 

\begin{figure}
\resizebox{\hsize}{!}{\includegraphics{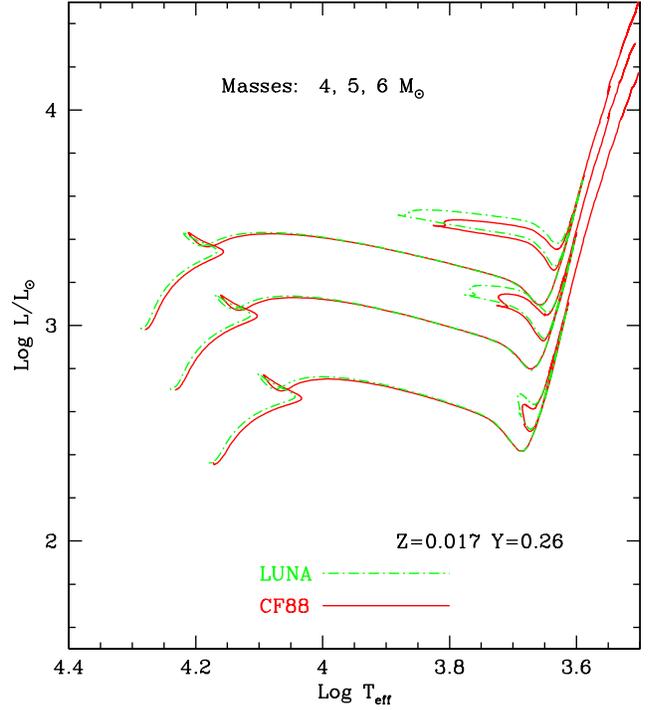}}
\caption{
Evolutionary tracks in the HR diagram for the composition $Z=0.017,
Y=0.26$ computed with the new LUNA rate for the 
$^{14}{\rm N}({\rm p},\gamma)^{15}{\rm O}$ nuclear reaction (dot-dashed line) 
and with the rate according to Caughlan \& Fowler (1988)(solid line). 
}

\label{LUNA_z017}
\end{figure} 

In our models, we tested the effect of measurements of the 
$^{14}{\rm N}({\rm p},\gamma)^{15}{\rm O}$ reaction rate by the LUNA 
experiment (Formicola et al. 2004) and from the NACRE compilation (Angulo et
al. 1999). The tracks obtained with the CF88 rate 
practically coincide with those obtained with the NACRE
rate, as shown by Imbriani et al. (2004). 
 The LUNA rate was simulated by simply multiplying that of
Caughlan \& Fowler  by a factor of 0.6 (LUNA Collaboration, Bemmerer et al.,
2006, Imbriani et al., 2005, Lemut et al., 2006).
The results are shown in Fig. 3 for Z=0.004 and Y=0.26, and in
Fig. 4 for Z=0.017 and Y=0.26 for the masses indicated in the
figures. Additional tracks for Z=0.008 were computed, but are not shown for
conciseness. From these tracks, we conclude that:  

\begin{itemize}
\item The MS becomes slightly bluer and slightly more luminous in models with 
a LUNA rate, as
a consequence of the H-burning lifetimes becoming shorter by about 1.5 \%. 
This is in general agreement with the results by Weiss et al. (2005).

\item The He-burning lifetimes do not change. He-burning loops have their
$T_{eff}$ extension practically unchanged for Z=0.004 and 0.008, whereas for 
Z=0.017 the loops become more extended with LUNA rates. This effect is 
surprising, and goes in the opposite sense to the results described in Weiss
et al. (2005). 
\end{itemize}

Overall, these changes in the tracks are very modest. Since the loops do not 
change at sub-solar metallicities, the comparison with Cepheid data are not 
affected.

\subsection{Changes in surface chemical composition}
\label{sec_chemical}

The surface chemical composition of the stellar models is affected by
two defined dredge-up events. The first occurs on
the first ascent of the RGB for all stellar models (except for 
the very-low mass ones which do not evolve beyond the main 
sequence). The second dredge-up is found after the core 
He-exhaustion, being remarkable only in models with 
$M\ga3.5~M_{\odot}$ and $M\la6~M_{\odot}$. For masses higher than
$6~M_{\odot}$, the second dredge-up occurs in concomitance with
the carbon ignition at the core, so that the exposition of He-enriched
material at the surface is too short-lived to be of interest, and
not followed in detail by the present calculations.
We provide tables with the 
surface chemical composition of H, $^3$He, $^4$He, and main CNO isotopes, 
before and after the first dredge-up, and after the second dredge-up.
As an example, Table 2 shows  the surface abundances for
the chemical composition Z=0.008 and Y=0.26.


\begin{table*}
\caption{Surface chemical composition (by mass fraction) of 
$[Z=0.008, Y=0.26]$ models.}
\label{tab_du}
\begin{tabular}{lllllllllll}
\noalign{\smallskip}\hline\noalign{\smallskip}
$M/M_{\odot}$ &	 H  &    $^3$He  &      $^4$He  &  $^{12}$C   &    $^{13}$C   &    $^{14}$N   &    $^{15}$N  &     $^{16}$O   &    $^{17}$O  &     $^{18}$O \\
\noalign{\smallskip}\hline\noalign{\smallskip}
\multicolumn{11}{l}{Initial:}   \\
  all   & 0.732 & 2.78$\:10^{-5}$ &  0.260 & 1.37$\:10^{-3}$ & 1.65$\:10^{-5}$ & 4.24$\:10^{-4}$ & 1.67$\:10^{-6}$ & 3.85$\:10^{-3}$ & 1.56$\:10^{-6}$ & 8.68$\:10^{-6}$ \\
\noalign{\smallskip}\hline\noalign{\smallskip}
\multicolumn{11}{l}{After the first dredge-up:} \\

  2.50 &  0.699 & 2.02$\:10^{-4}$ & 0.292 & 8.50$\:10^{-4}$ & 4.40$\:10^{-5}$ & 1.32$\:10^{-3}$ & 8.18$\:10^{-7}$ & 3.49$\:10^{-3}$ & 1.05$\:10^{-5}$ & 6.25$\:10^{-6}$ \\  
  3.00 &  0.700 & 1.42$\:10^{-4}$ & 0.292 & 8.48$\:10^{-4}$ & 4.45$\:10^{-5}$ & 1.36$\:10^{-3}$ & 8.11$\:10^{-7}$ & 3.44$\:10^{-3}$ & 7.60$\:10^{-6}$ & 6.24$\:10^{-6}$ \\  
  3.50 &  0.702 & 1.07$\:10^{-4}$ & 0.290 & 8.50$\:10^{-4}$ & 4.46$\:10^{-5}$ & 1.37$\:10^{-3}$ & 8.10$\:10^{-7}$ & 3.42$\:10^{-3}$ & 6.13$\:10^{-6}$ & 6.24$\:10^{-6}$ \\  
  4.00 &  0.704 & 8.61$\:10^{-5}$ & 0.287 & 8.62$\:10^{-4}$ & 4.54$\:10^{-5}$ & 1.35$\:10^{-3}$ & 8.12$\:10^{-7}$ & 3.43$\:10^{-3}$ & 4.89$\:10^{-6}$ & 6.32$\:10^{-6}$ \\  
  4.50 &  0.707 & 7.16$\:10^{-5}$ & 0.285 & 8.72$\:10^{-4}$ & 4.49$\:10^{-5}$ & 1.34$\:10^{-3}$ & 8.16$\:10^{-7}$ & 3.43$\:10^{-3}$ & 4.23$\:10^{-6}$ & 6.33$\:10^{-6}$ \\  
  5.00 &  0.708 & 6.04$\:10^{-5}$ & 0.284 & 8.85$\:10^{-4}$ & 4.52$\:10^{-5}$ & 1.33$\:10^{-3}$ & 8.26$\:10^{-7}$ & 3.43$\:10^{-3}$ & 4.05$\:10^{-6}$ & 6.37$\:10^{-6}$ \\  
  6.00 &  0.710 & 4.70$\:10^{-5}$ & 0.282 & 8.82$\:10^{-4}$ & 4.74$\:10^{-5}$ & 1.31$\:10^{-3}$ & 8.09$\:10^{-7}$ & 3.45$\:10^{-3}$ & 3.46$\:10^{-6}$ & 6.42$\:10^{-6}$ \\  
  7.00 &  0.711 & 3.89$\:10^{-5}$ & 0.281 & 8.81$\:10^{-4}$ & 4.76$\:10^{-5}$ & 1.32$\:10^{-3}$ & 8.00$\:10^{-7}$ & 3.45$\:10^{-3}$ & 3.06$\:10^{-6}$ & 6.40$\:10^{-6}$ \\  
  8.00 &  0.711 & 3.35$\:10^{-5}$ & 0.281 & 8.77$\:10^{-4}$ & 4.84$\:10^{-5}$ & 1.33$\:10^{-3}$ & 7.90$\:10^{-7}$ & 3.44$\:10^{-3}$ & 3.02$\:10^{-6}$ & 6.37$\:10^{-6}$ \\  
  10.0 &  0.711 & 2.73$\:10^{-5}$ & 0.281 & 8.75$\:10^{-4}$ & 4.91$\:10^{-5}$ & 1.33$\:10^{-3}$ & 7.75$\:10^{-7}$ & 3.44$\:10^{-3}$ & 3.10$\:10^{-6}$ & 6.33$\:10^{-6}$ \\  
  12.0 &  0.703 & 2.34$\:10^{-5}$ & 0.289 & 8.65$\:10^{-4}$ & 5.03$\:10^{-5}$ & 1.42$\:10^{-3}$ & 7.56$\:10^{-7}$ & 3.34$\:10^{-3}$ & 2.71$\:10^{-6}$ & 6.19$\:10^{-6}$ \\  
  15.0 &  0.683 & 1.95$\:10^{-5}$ & 0.309 & 8.30$\:10^{-4}$ & 5.01$\:10^{-5}$ & 1.64$\:10^{-3}$ & 7.16$\:10^{-7}$ & 3.15$\:10^{-3}$ & 2.24$\:10^{-6}$ & 5.86$\:10^{-6}$ \\  
  20.0 &  0.731 & 2.54$\:10^{-5}$ & 0.261 & 1.23$\:10^{-3}$ & 6.69$\:10^{-5}$ & 5.49$\:10^{-4}$ & 1.00$\:10^{-6}$ & 3.83$\:10^{-3}$ & 1.60$\:10^{-6}$ & 8.42$\:10^{-6}$ \\  
\noalign{\smallskip}\hline\noalign{\smallskip}
\multicolumn{11}{l}{After the second dredge-up:} \\

  3.50 &  0.693 & 1.03$\:10^{-4}$ & 0.298 & 8.22$\:10^{-4}$ & 4.46$\:10^{-5}$ & 1.45$\:10^{-3}$ & 7.77$\:10^{-7}$ & 3.37$\:10^{-3}$ & 6.98$\:10^{-6}$ & 6.06$\:10^{-6}$ \\  
  4.00 &  0.675 & 8.00$\:10^{-5}$ & 0.317 & 8.08$\:10^{-4}$ & 4.48$\:10^{-5}$ & 1.55$\:10^{-3}$ & 7.56$\:10^{-7}$ & 3.28$\:10^{-3}$ & 5.60$\:10^{-6}$ & 5.93$\:10^{-6}$ \\  
  4.50 &  0.658 & 6.44$\:10^{-5}$ & 0.334 & 7.95$\:10^{-4}$ & 4.40$\:10^{-5}$ & 1.64$\:10^{-3}$ & 7.40$\:10^{-7}$ & 3.19$\:10^{-3}$ & 4.76$\:10^{-6}$ & 5.77$\:10^{-6}$ \\  
  5.00 &  0.644 & 5.27$\:10^{-5}$ & 0.348 & 7.85$\:10^{-4}$ & 4.39$\:10^{-5}$ & 1.72$\:10^{-3}$ & 7.29$\:10^{-7}$ & 3.12$\:10^{-3}$ & 4.45$\:10^{-6}$ & 5.63$\:10^{-6}$ \\  

\noalign{\smallskip}\hline\noalign{\smallskip}
\end{tabular}
\end{table*}

\begin{figure}
 \resizebox{\hsize}{!}{\includegraphics{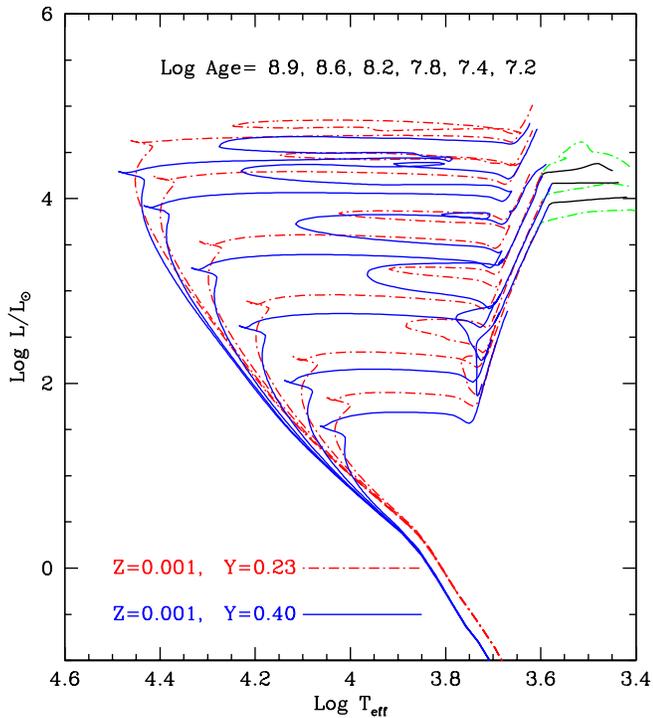}}
\caption{Comparison of isochrones for the same Z=0.001 and different helium 
content. Blue solid lines correspond to Y=0.40 (the related AGB is in black),
and red dot-dashed ones to Y=0.23 (the related AGB is in green).
 }
\label{isozy1y2}
\end{figure}

\section{Isochrones}
\label{sec_isochrones}

From the tracks presented in this paper, we constructed isochrones
as in Paper I, with the algorithm of ``equivalent evolutionary points'' 
used in Bertelli et al.\ (1994) and Girardi et al. (2000).   
The initial point of each isochrone is the $0.15 M_{\odot}$ model in the 
lower main sequence. The terminal stage of the isochrone is 
the tip of the TP-AGB or the beginning of carbon burning. 
We note that there are usually small 
differences in the logarithm of the effective temperature between the last
model of the track at the beginning of the TP-AGB and the starting point
of the synthetic TP-AGB model of the corresponding mass. Constructing the
isochrones, we removed these small discontinuities with a suitable shift. 

A comparison of isochrones with the same metallicity ($Z=0.001$) and 
different helium content ($Y=0.23$ and $Y=0.40$) can be found in 
Fig. 5 showing the range of variations in luminosity and
temperature for some isochrones ($\log (Age/{\rm yr}) = 8.9, 8.6, 8.2, 7.8, 
7.4, 7.2$).
The extension to the end of the TP-AGB phase is also shown for each isochrone.
An increase in the helium content at the same metallicity in stellar 
models causes a decrease in the mean opacity and an increase in the
mean molecular weight of the envelope (Vemury and Stothers, 1978), and
in turn higher luminosities, hotter effective temperatures and shorter
hydrogen and helium lifetimes of stellar models.
In apparently conflict with the previous statement about 
higher luminosity for a helium increase in evolutionary tracks, the evolved
portion of the isochrones with lower helium are more luminous, as shown in 
Fig. 5 where we plot isochrones with Z=0.001 for Y=0.23 and 
Y=0.40.  This effect is related to the 
interplay between the increase in luminosity and the decrease in lifetime 
of stellar models with higher helium content (at the same mass and 
metallicity).  

\subsection{Reliability of the interpolation}

The program YZVAR is used  to both obtain isochrones and simulate
stellar populations. The method of
interpolation has been described in detail in Paper I.
In the database, there are 37 sets of stellar tracks covering a large region of
the $Z-Y$ plain and users can  obtain either isochrones for whichever $Z-Y$
combination inside the provided range, or stellar populations with the required
Y(Z) enrichment law. 

Most of the problems about the reliability of the interpolation originate 
from the extension of the loops during the central He-burning, which varies
significantly from low to high metal content for the considered range of mass.
For example, if we consider the two subsequent values of $Z$ of our grid, 
$Z=0.017$ and $Z=0.040$, we have extended loops for the first one (starting 
from a mass that increases with helium content $Y$). No loops are present 
for the upper mass $20 M_{\odot}$ for $Y > 0.26$. 
In the case of $Z=0.040$, for all $Y$ values and masses, no loops are 
present, as shown in Fig. 2. 
In this case, we cannot expect the extension of the loops to have  a linear 
behaviour  
as a function of the metal content in the interpolation for intermediate 
chemical compositions. We analysed the interpolation for $Z=0.0285$ and two 
intermediate cases, namely $Y=0.28$ and  $Y=0.37$. 

In the case of $Z=0.0285, Y=0.28$ the loops obtained by interpolation are 
significantly less extended and/or less luminous than those of tracks 
computed for the same
chemical composition at $7$ and $10 M_{\odot}$. In contrast, the 
interpolated tracks of $15$ and $20 M_{\odot}$ exhibit significant loops, while
in the computed ones there are no loops.
In Fig. 6 the solid green lines represent the 
computed tracks, and the red dotted lines represent the interpolated tracks.

 To improve the interpolation for this range of metal content, we computed new 
tracks with $Z=0.030$, $Y=0.26, 0.30, 0.34,$ and $ 0.40$ for 
$M > 4.5 M_{\odot}$.
In this, way the results of the interpolation become far more reliable, as  
evident in Fig. 6, where the blue line shows the result of the 
interpolation after taking into account this new set for masses greater than 
$4.5 M_{\odot}$. The comparison of the lifetimes of the computed and the 
interpolated tracks infers differences of lower than $3 \%$.
The interpolation is also superior in the case of  $Z=0.0285$ and 
$Y=0.37$, when taking into account the new tracks computed for the intermediate
metal content  $Z=0.030$.   

Another critical region of the interpolation is at
low metallicities (of $Z$ between 0.0004 and 0.001 for high values of $Y$)
owing to the different morphological behaviour of the involved tracks.
A computed set of tracks ($5, 7, 10, 15, 20 M_{\odot}$) with $Z=0.0007$ and 
$Y=0.35$ was compared with a corresponding interpolated set and the 
results are shown in Fig. 7, where the green lines represent the 
computed tracks and the red dotted lines the interpolated ones. There are
some differences in the red region, which is reached for the first time after 
the central
hydrogen exhaustion for the $7$ and $10 M_{\odot}$. If we check the lifetime
of the phases at variance with the computed tracks (see Fig. 8), the 
differences are negligible with respect to the total He-burning lifetime.     

In all other cases, the comparisons are satisfactory.

\begin{figure}
 \resizebox{\hsize}{!}{\includegraphics{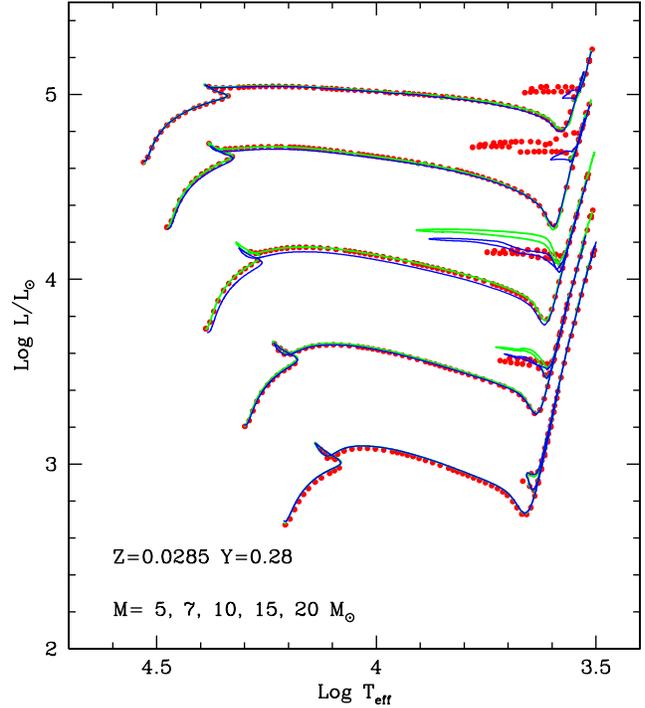}}
\caption{  
Comparison between interpolated and computed tracks with $Z=0.0285, Y=0.28$.
The red dotted lines are the interpolated tracks between $Z=0.017$ and
$Z=0.040$, while the green solid lines are the computed ones for the specific 
Z.
The blue lines are the interpolated tracks taking into account the new set
computed for $Z=0.030$ and various $Y$. Taking into account 
the new set, the interpolation is in closer agreement with the actual computed 
tracks. 
}
\label{int01}
\end{figure}

\begin{figure}
 \resizebox{\hsize}{!}{\includegraphics{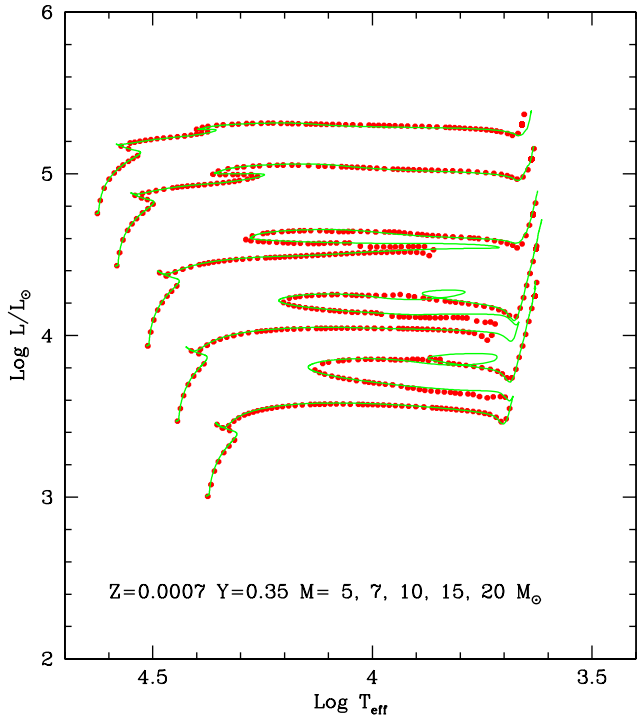}}
\caption{  
Comparison between interpolated and computed tracks with $Z=0.0007, Y=0.35$.
The red dotted lines are the interpolated tracks for masses between $5$ and 
$20 M_{\odot}$ and the green solid lines the computed lines for the specific
Z.
 }
\label{int02}
\end{figure}

\begin{figure}
 \resizebox{\hsize}{!}{\includegraphics{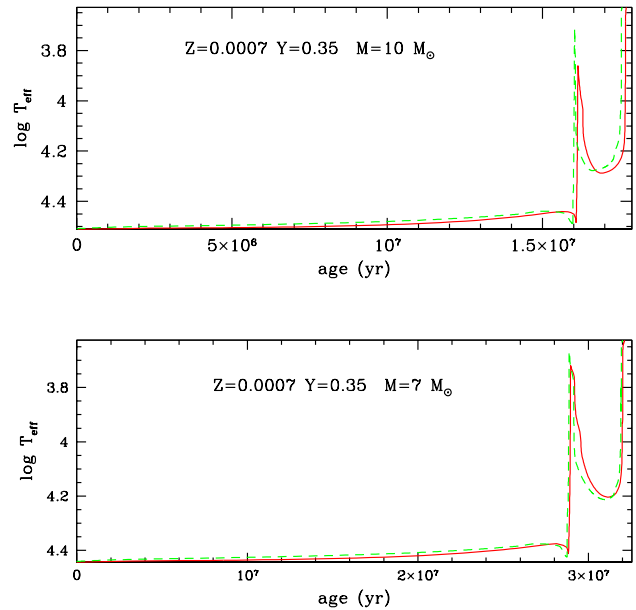}}
\caption{  
Differences between the lifetime spent by the interpolated track (red 
solid line) and that of the computed track (green dashed line) are 
insignificant. The chemical composition is as in Fig. 7. 
 }
\label{int03}
\end{figure}

\subsection{Bolometric corrections}

The present isochrones are provided in the Johnson-Cousins-Glass
system as defined by Bessell (1990) and Bessell \& Brett (1988), in the Vegamag
systems of ACS onboard HST (cf. Sirianni et al. 2005), and in the SDSS 
system.  The formalism that we follow to derive
bolometric corrections in these systems is described in Girardi et
al. (2002). The definition of zeropoints was revised as described
 in a paper by Girardi et al. (2008), a description that is not repeated here.

Suffice it to say that the bolometric correction tables are based
on an updated and extended library of stellar spectral fluxes. The core
of the library now consists of the ``ODFNEW'' ATLAS9 spectral fluxes
from Castelli \& Kurucz (2003), for $T_{\rm eff}$ of between 3500 and
50000~K, $\log g$ of between $-2$ and $5$, and scaled-solar metallicities
[M/H] of between -2.5 and +0.5. This library is extended to
 high $T_{\rm eff}$ by using pure black-body spectra.
For lower $T_{\rm eff}$, the library is completed
with  spectral fluxes for M, L, and T dwarfs from Allard et
al. (2000), M giants from Fluks et al. (1994), and finally  C star
spectra from Loidl et al. (2001).  Details about the implementation of
this library, and in particular about the C star spectra, are provided
in Marigo et al. (2008) and Girardi et al. (2008).

We note that in the isochrones we apply the
bolometric corrections derived from this library without making any
correction for the enhanced He content. As demonstrated in Girardi et
al. (2007), for a given metal content, an enhancement of He 
as high as
$\Delta Y=0.1$ produces changes in the bolometric corrections of only a
few thousandths of magnitude. Just in some particular situations,
for instance at low $T_{\rm eff}$ and for blue passbands, the
He-enhancement can produce more sizeable effects on BCs; these situations,
however, correspond to cases where the emitted stellar flux would
in all cases be very small, and therefore are of little interest.

\subsection{Description of isochrone tables}
\label{sec_tableisoc}
                        
Complete tables with the isochrones can be obtained through our web site
\footnote{ http://stev.oapd.inaf.it/YZVAR} interactive services.

The user must select the chemical composition, the $\Delta\log t$ spacing
between contiguous isochrones, and the photometric system.  
 The following  photometric systems are available:

\begin{itemize}
\item UBVRIJHK absolute magnitudes in the Johnson-Cousins-Glass photometric 
system 

\item ACS-WFC photometric system (bolometric corrections  from Girardi et al. 
2008). In the tables, the absolute magnitudes for the following 12 filters are 
listed:  F435W, F475W, F550M, 
F555W, F606W, F625W, F658N, F660N, F775W, F814W, F850LP, F892N 

\item ACS-HRC photometric system (see Girardi et al. 2008). The absolute
magnitudes listed in the tables are for the following 16 filters: F220W, F250W,
F330W, F344N,  F435W, F475W, F550M, F555W, F606W, F625W, F658N, F660N, F775W, 
F814W, F850LP, F892N. 

\end{itemize}

In the case of the Johnson-Cousins-Glass system,
the isochrone tables contain the following information: 

        \begin{description}
	\item \verb$1. logAge$: logarithm of the age in years;
	\item \verb$2. log(L/Lo)$: logarithm of surface luminosity (in solar units);
	\item \verb$3. logTef$: logarithm of effective temperature (in K);
	\item \verb$4. logG$: logarithm of surface gravity (in cgs units);
        \item \verb$5. Mi$: initial mass in solar masses;
        \item \verb$6. Mcur$: actual stellar mass in solar masses;  
        \item \verb$7. FLUM$: indefinite integral over the initial mass M of 
the Salpeter initial mass function by number;
        \item \verb$8. - 15.$: UBVRIJHK absolute magnitudes in the 
Johnson-Cousins-Glass system; 
        \item \verb$16. C/O $  the surface C/O ratio;
        \item \verb$17. C.P.$: index marking the presence of a characteristic 
point, when different from zero.
        \end{description}

The isochrone tables for the ACS-WFC photometric system contain the same 
information in the first 7 columns, then the absolute
magnitudes for this system from Cols. 8 to 19, and the last two columns
are the same as in the previous case.
For the ACS-HRC photometric system from Cols. 8 to 23 the absolute
magnitudes for the relative filters are presented. 

 We note that the
initial mass is the useful quantity for population synthesis calculations,
since together with the initial mass function it determines the relative 
number of stars in different sections of the isochrones.
 In  Col. 7,  the 
indefinite integral evaluated over the initial mass $M$ of the initial mass 
function (IMF) by number, i.e.,
	\begin{equation}
\mbox{\sc flum} = \int\phi(M) \diff M
	\end{equation}
is presented, for the case of the Salpeter IMF, $\phi(M)=AM^{-\alpha}$, 
were $\alpha=2.35$. Setting $A=1$, 
{\sc flum} is simply given by {\sc flum}$ = M^{1-\alpha}/(1-\alpha)$.
This is a useful quantity since the difference between any two 
values of {\sc flum} is proportional to the number of stars located in 
the corresponding isochrone interval. 
For example, if we consider an old isochrone fitting the observations of a 
globular cluster, and along the RGB we consider three points identifying
two intervals ($1-2$ and $2-3$), the ratio of the differences of $\mbox{\sc flum}$ at each interval borders
gives exactly the expected ratio of the number of stars in the corresponding
region of the observed globular cluster,  as shown in the  relation:

 $(\mbox{\sc flum}_3 -\mbox{\sc flum}_2) /(\mbox{\sc flum}_2 - \mbox{\sc flum}_1)  = {\rm N}^*_{2-3} / {\rm N}^*_{1-2}$,

\noindent where the terms have their usual meaning.

We remark that the 
Salpeter (1955) IMF is not valid for masses lower than about $1 M_{\odot}$.
 The user can easily derive {\sc flum} relations for alternative choices of
the IMF, by using the values of the initial mass that we present in Col. 5 of 
the isochrone tables.
Finally, the last column may signal the presence of a characteristic
evolutionary point. Values of 1 refer to points from the ZAMS to the early
AGB phase, and values of 2 represent the TP-AGB phase. Along the TP-AGB, we 
mark its beginning, its end, and the point where C/O increases above unity,
corresponding to the transition from M to carbon stars, if present. The 
presence of only one characteristic point of type 2 indicates a very short 
TP-AGB phase.  

\section{Cepheids}

From an evolutionary perspective, Cepheids are post-red giant
stars crossing the instability strip on so-called blue loops following the
onset of core He-burning. Obviously the extension of the loop and the 
distribution of the lifetime along it determines the appearance and the 
distribution of the  Cepheids in the HR diagram. Many factors can modify the 
loops, including all those able to change the relative contribution
to the total luminosity from the H-burning shell and the central He-burning
core. In this respect, a crucial effect is produced by the physical treatment
of the central and envelope convection and related uncertainties. 
For the intermediate-mass stars, appreciable core overshooting during the 
main sequence 
causes a smaller loop, while envelope overshooting  during 
the RGB phase tends to favour more extended loops. 
The parameters characterizing the core and envelope overshoot in
Padova models were calibrated with observational constraints (Alongi 
et al 1991,1993 and ref therein).   
In  Figs 9, 10, and 11, we present 
tracks for $2.5, 3, 4, 5, 6, 7, 8,$ and $ 10 M_{\odot}$ 
for values of the metallicity  $Z=0.002$, $Z=0.004$, and $Z=0.008$.  
To compare the extension of the loops,
in the same figures we plotted the corresponding tracks computed 
by the Teramo group (TE04, Pietrinferni et al 2004). These TE04 stellar models
have  an upper mass limit of $10 M_{\odot}$ in their database.
The main-sequence phases and the  contraction towards the Hayashi limit 
are very similar for the two evolutionary grids. Our red giant phase
appears slightly redder because of a different
choice of mixing length parameter. The loops have different extensions,
but this is  normal because of the sensitivity of the loops to many of the
input parameters. 

\begin{figure}
 \resizebox{\hsize}{!}{\includegraphics{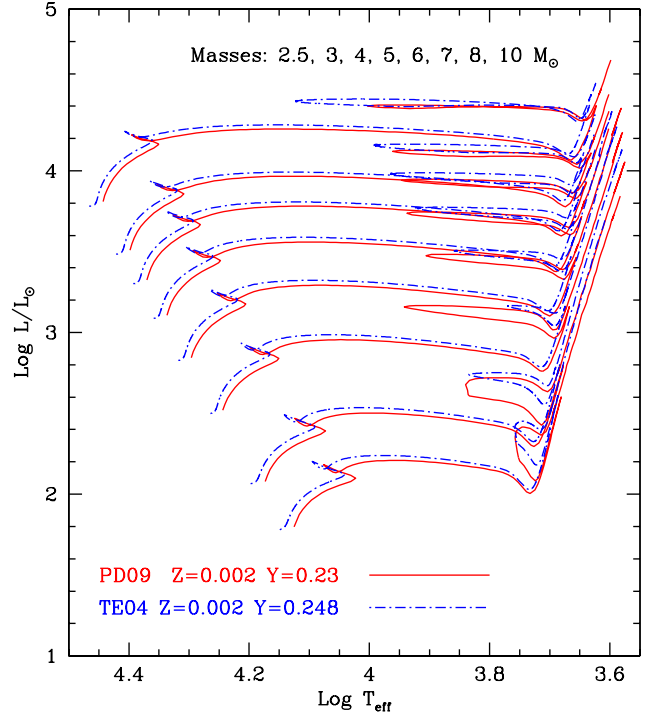}}
\caption{  
PD09 evolutionary tracks with $Z=0.002, Y=0.23$ for masses between $2.5$ and 
$10 M_{\odot}$ (solid line). TE04 models with $Z=0.002, Y=0.248$ (dot-dashed
line) for the same mass range.  
 }
\label{ceph1a}
\end{figure}

\begin{figure}
 \resizebox{\hsize}{!}{\includegraphics{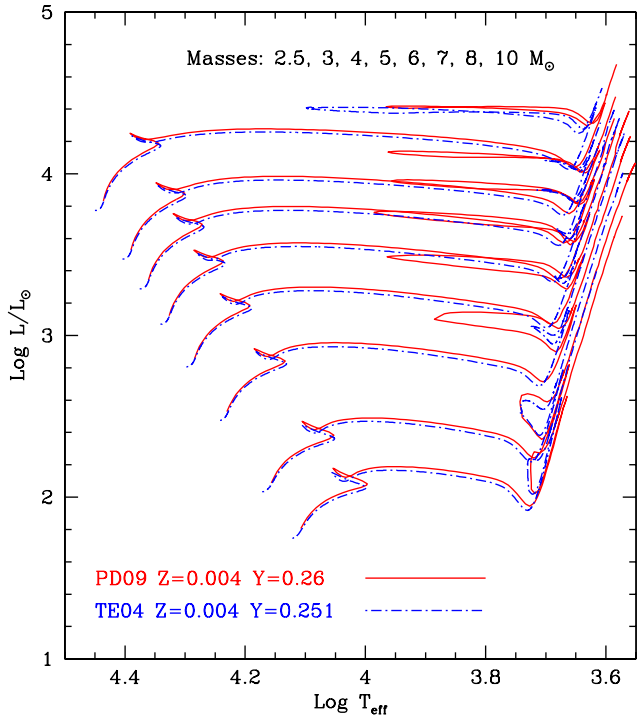}}
\caption{  
PD09 evolutionary tracks with $Z=0.004, Y=0.26$ for masses between $2.5$ and 
$10 M_{\odot}$ (solid line). TE04 models with $Z=0.004, Y=0.251$ (dot-dashed
line) for the same mass range. It is evident in TE04 models the absence of 
loops for masses lower than $10 M_{\odot}$.
 }
\label{ceph1b}
\end{figure}

\begin{figure}
 \resizebox{\hsize}{!}{\includegraphics{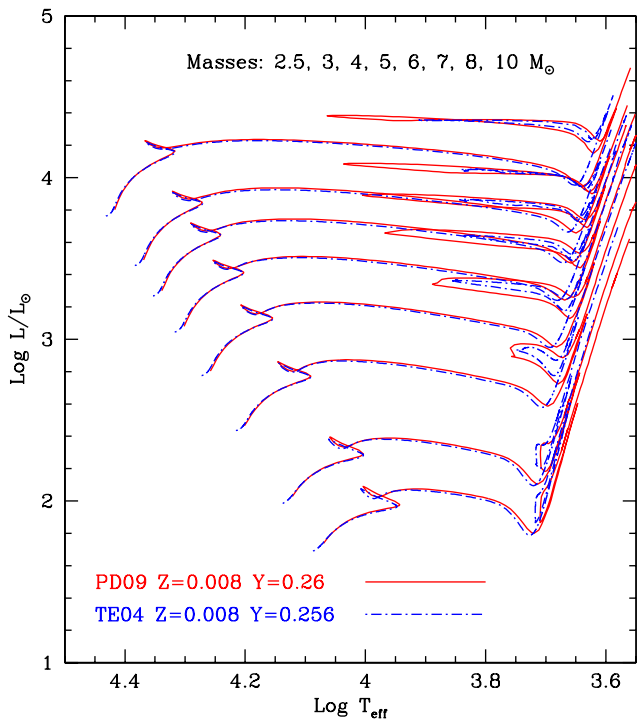}}
\caption{  
PD09 evolutionary tracks with $Z=0.008, Y=0.26$ for masses between $2.5$ and 
$10 M_{\odot}$ (solid line). TE04 models with $Z=0.008, Y=0.256$ (dot-dashed
line) for the same mass range.  
 }
\label{ceph1c}
\end{figure}

The only true and puzzling difference is that related to
 the absence of loops in TE04 tracks at lower mass in the case $Z=0.004$.  
Our tracks illustrate a typical behaviour, in the sense that the minimum mass 
exhibiting a significant loop (able to intercept the instability strip 
of Cepheids) tends to increase with the metal content $Z$ (and with the 
 helium content $Y$ at fixed metallicity). 
This prediction is 
confirmed by all the chemical compositions considered in our computations. 

In contrast, the case $Z=0.004$ of TE04 appears to contradict this pattern,
since only the $10 M_{\odot}$ model exhibits an extended loop, while for 
$Z=0.002$ and $Z=0.008$ loops are present at lower masses.
This is at variance with the period (mass) distribution of Cepheids in the
SMC, for which the metal content $Z=0.004$ is usually adopted.

 A similar difference was already shown by evolutionary tracks
of intermediate-mass stars with no convective overshooting discussed in 
Castellani et al (1990). They explained the decrease in the extension of 
the blue loop for masses around 5 $M_{\odot}$ and the intermediate metal 
content $Z=0.006$ and $Y=0.23$, connected with the occurrence of the 
first dredge-up. It is not obvious why this feature is not present for
higher $Y$ values and/or different $Z$, as would be expected.

\subsection{Comparison with the observations: SMC, LMC, Milky Way}

Cordier et al (2003) explored the distribution 
of SMC Cepheids in the HR diagram in relation to the extension 
of the loops of the evolutionary tracks. They showed that the blue loops 
are sufficiently extended only if the metallicity is substantially lower
than  the usually adopted value for the SMC. They found that tracks with 
metallicity
$Z=0.001$ correctly reproduce the boundary of the Cepheids towards lower
luminosity. According to Cordier et al (2003), Cepheids populating 
the faint bottom
of the instability strip are likely to be metal-poor stars, while
variables distributed throughout the
instability strip should belong to populations of various metal contents. 

Romaniello et al (2008) derived direct measurements of the iron
abundances of Galactic and Magellanic Cepheids from FEROS and UVES 
high-resolution and high signal-to noise spectra. The mean iron abundance 
($[$Fe/H$]$) was found to be
about solar for the Galactic sample, with a range of values 
between $-0.18$ dex and $+0.25$ dex. For the LMC sample, the mean value is 
about $-0.33$ dex with a range of values between $-0.62$ dex and $-0.10$ 
dex. For the SMC sample, the mean value is about $-0.75$ dex with
a range of values between $-0.87$ and $-0.63$ dex.

If the behaviour of the loops in the evolutionary models is
correct (i.e., the increase of the minimum mass that intercepts the 
instability strip with $Z$ and $Y$), this means that the lower 
limit to data for Cepheids in the HR diagram is determined by the stars of lower
metal content $Z$. In Figs 12, 13, and 14 we checked whether our models
can reproduce the lower boundaries of the instability strip compatibly with 
the dispersion in $Z$ found by Romaniello et al (2008) for SMC, LMC, and MW 
respectively.   

\begin{figure}
 \resizebox{\hsize}{!}{\includegraphics{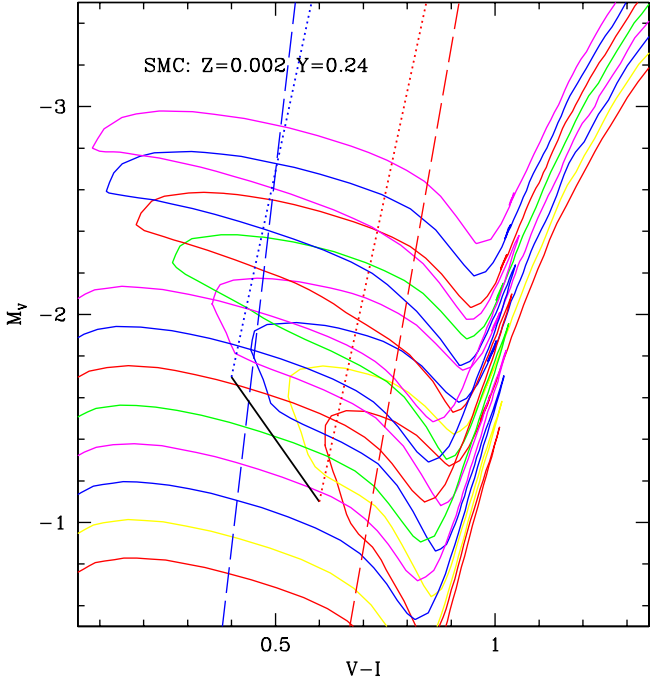}}
\caption{  
PD09 isochrones with $Z=0.002, Y=0.24$ for ages in the range $8.30 \le 
\log(t/{\rm yr}) \le 8.65$. Long-dashed lines represent the blue and red edges 
of the theoretical instability strip (from Caputo et al. 2004). The dotted 
lines are the observed boundaries,
and the black line represents the bottom of the instability strip
from observations of SMC Cepheids (as in Fig. 7 by Sandage et al. 2009). 
 }
\label{cepsmc}
\end{figure}

\begin{figure}
 \resizebox{\hsize}{!}{\includegraphics{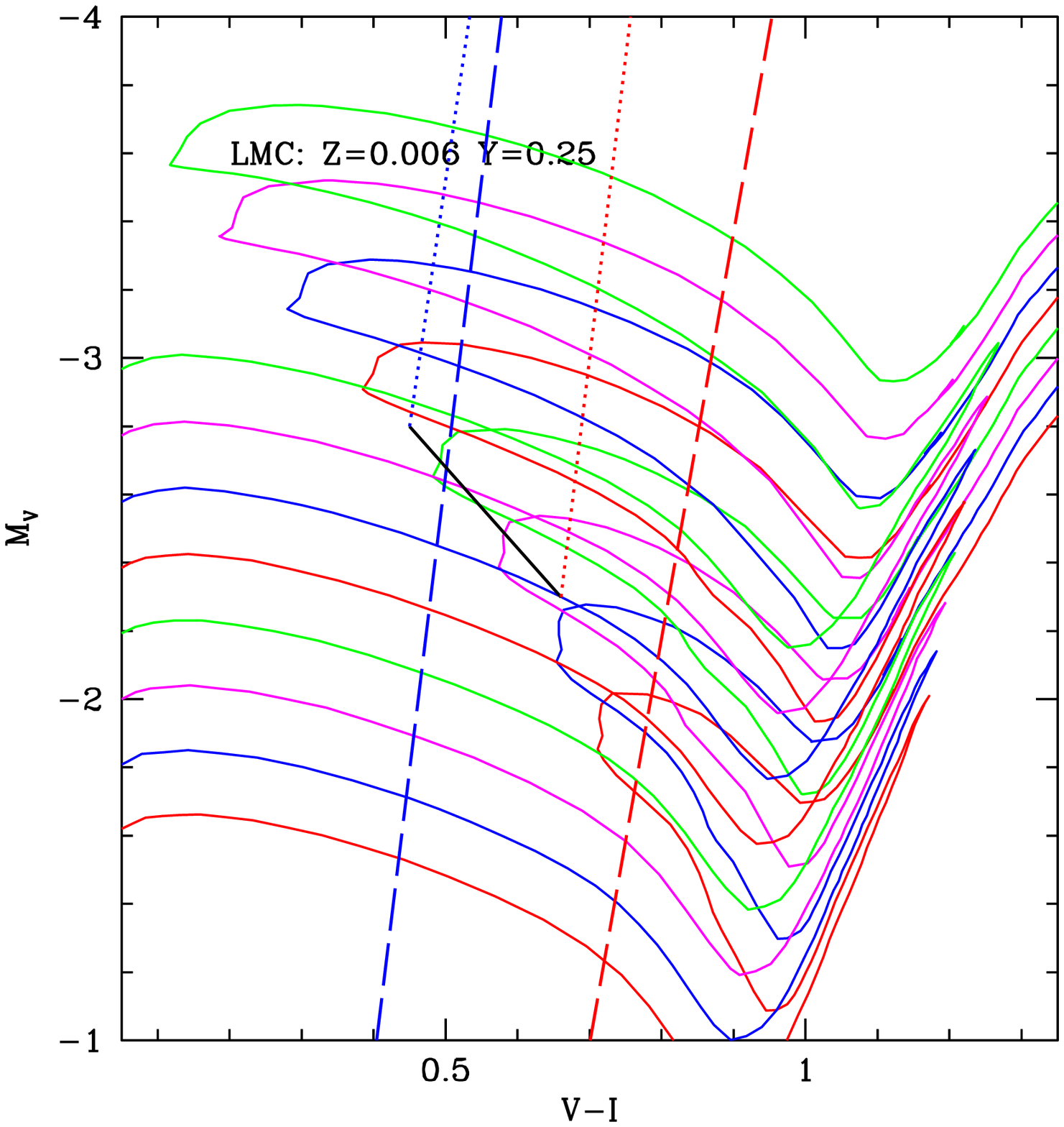}}
\caption{  
PD09 isochrones with $Z=0.006, Y=0.25$ for ages in the range $8.05 \le 
\log(t/{\rm yr}) \le 8.40$. Long-dashed lines represent the blue and red edges 
of the theoretical instability strip (from Caputo et al. 2004). The dotted 
lines are the observed boundaries,
and the black line represents the bottom of the instability strip
from observations of LMC Cepheids (from Fig. 8 in Sandage et al. 2004). 
 }
\label{ceplmc}
\end{figure}

\begin{figure}
 \resizebox{\hsize}{!}{\includegraphics{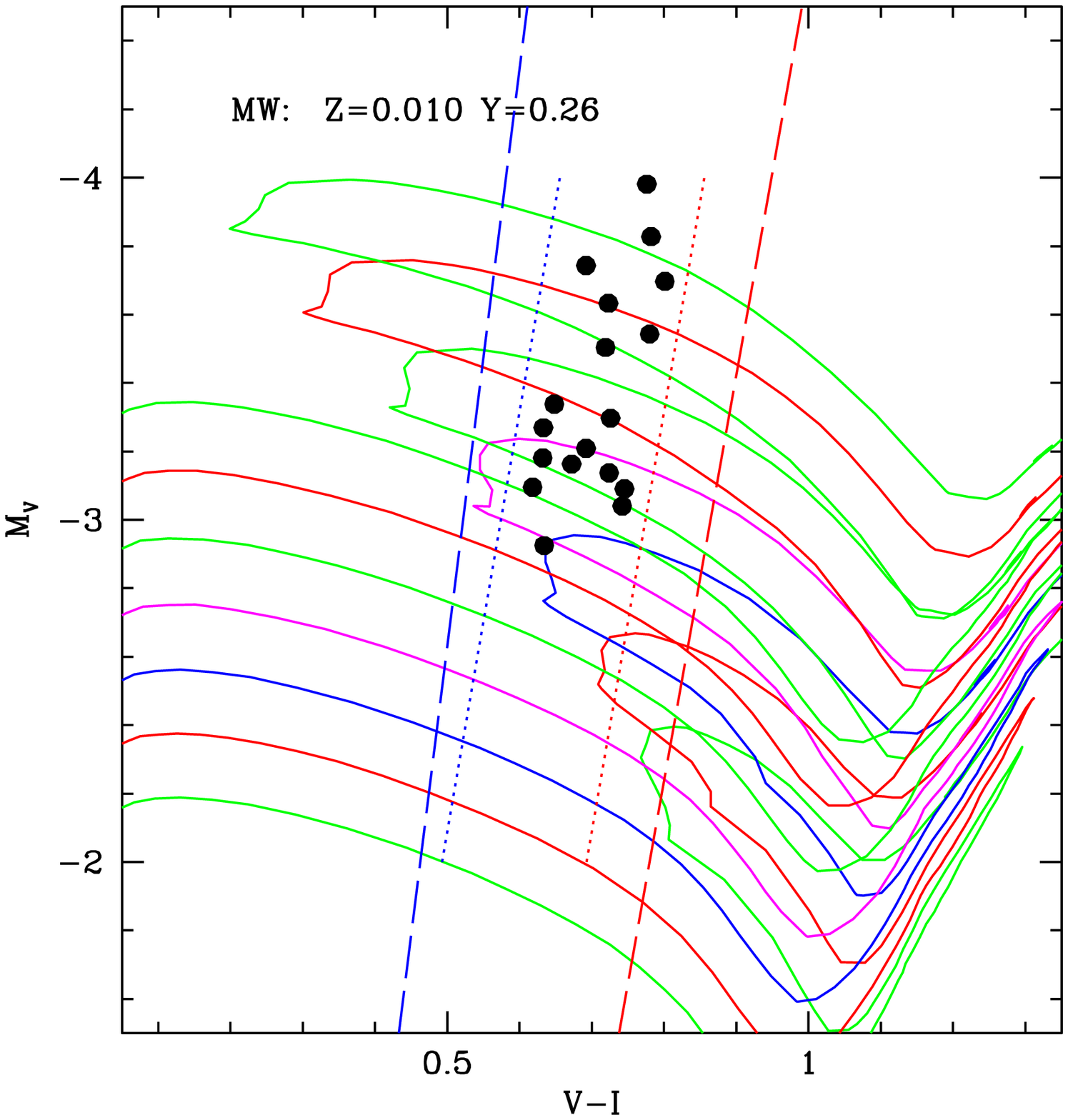}}
\caption{  
PD09 isochrones with $Z=0.010, Y=0.26$ for ages in the range $7.95 \le 
\log(t/{\rm yr}) \le 8.25$. Long-dashed lines represent the blue and red edges 
of the theoretical instability strip (from Caputo et al. 2004). The dotted 
lines are the observed boundaries, and the Cepheids in open clusters and 
associations are plotted (from Tammann et al. 2003).
 }
\label{cepmw}
\end{figure}

In Fig. 12 we show the loops of isochrones with ages in the range 
$8.30 \le \log(t/{\rm yr}) \le 8.65$ (the separation between the isochrones is 
$\Delta \log(t/{\rm yr})=0.05$) and 
chemical composition $Z=0.002$, $Y=0.24$. The blue and red
long dashed lines represent, respectively, the blue and red edges of the
theoretical instability strip as obtained by Eqs. 1a and 2a in 
Caputo et al (2004),
where we adopt the mass-luminosity relation calibrated in our models. 
The black thick line represents the lower boundary of the instability strip
from  Fig. 7 of the paper on the SMC Cepheids by Sandage et al (2009). 
The blue and red dotted lines are the experimental 
boundaries of the strip  shown in the same Fig. 7 of Sandage et al.  
We note that:
\begin{itemize}
\item the observed instability strip is narrower than the theoretical one;
\item the slope of the lower boundary for observed stars coincides with the 
slope of the faint envelope of the loops; 
\item the masses of the stars developing loops close to the
   observed lower boundary are in the range $2.5 -  3 M_{\odot}$;
\item  the observed lower limit could be recovered by models by  adopting a
slightly lower metal content than the assumed $Z=0.002$ (as given by 
Romaniello et al. 2008).
\end{itemize}
 
In Fig. 13, we plot the results for the Cepheids of the LMC. In this 
case, the chemical composition adopted for the set of isochrones is $Z=0.006$, 
$Y=0.25$.
The bottom of the instability strip was taken from Fig. 8 of the 
paper on the LMC Cepheids by Sandage et al (2004). The blue and red
dotted lines correspond to the blue and red boundaries of the instability
strip, such that the strip contains $ \sim 90 \% $ of the Cepheids (Fig. 8 of
Sandage et al 2004). The blue and red long-dashed  lines are the boundaries 
of the theoretical strip as represented by the formulae of Caputo et al (2004) 
for the appropriate chemical composition. We point out that
\begin{itemize}
\item  in the case of the LMC, the observed strip is also narrower than the 
theoretical one (the red edges differ significantly in position 
but not in slope); 
\item the ages of the loops shown  are in the interval 
$8.05 \le \log(t/{\rm yr})
\le 8.40$ (with $\Delta \log(t/{\rm yr})=0.05$) and the masses of models with
loops close to the observed bottom are in  the range  $3.7 - 
4.4 M_{\odot}$;   
\item the chemical composition adopted in our analysis ($Z=0.006$) that can 
reproduce the  observations  is $\sim 40 \%$ higher than the lower 
limit indicated by Romaniello et al. ($Z=0.004$) and $\sim 28 \%$ lower than 
the mean value ($Z=0.008$), which is usually adopted for the LMC.  
\end{itemize}

 In the case of the Galaxy, we  considered the calibration of Cepheids in 
open clusters and associations listed in Table 3 of the paper by Tammann et al 
(2003).
In Fig. 14 we  plotted this sample of Cepheids over  a set of  
isochrones with
metal content $Z=0.010$ and $Y=0.26$ (this value of the metallicity corresponds
to the lower limit of metallicity for the Galactic Cepheids  as given in 
Romaniello et al. 2008 ). The observed instability strip is reproduced using  
Eqs. (5) and (16), and Fig. 15 of Tammann et al (2003).   
Because of the small number of objects, it is impossible
to draw the lower boundary of the observed strip  as done for the LMC and the 
SMC. The only   
conclusion is that the value $Z=0.010$ is compatible with the data in the sense
 that the observed object with the highest magnitude lies very close to the  
faint envelope detected by the loops  of the isochrones shown in 
Fig. 14.

\subsection{Mass discrepancy}

A long-standing open question  is given by the
discrepancy between the pulsation and the evolutionary mass of
the Cepheids (for a brief review see Keller, 2008).
Caputo et al (2005) considered 34 Cepheids in the Milky Way
with solar-like metal content previously studied by Storm et al (2004).
They determined the pulsation mass $M_{\rm p}$ from the predicted PLC relations
and the evolutionary one $M_{\rm e}$ from the mass-period-luminosity (MPL) or 
the mass-color-luminosity (MCL) relations adopting
the canonical\footnote{By this term, we mean without overshooting 
from the convective core} relation mass-luminosity (ML). 
The comparison between the two determinations of the mass showed that
the $M_{\rm p}/M_{\rm e}$ ratio is correlated with the Cepheid period, ranging 
from $\sim 0.8$ at $\log(P/{\rm days})=0.5$ to $\sim 1$ at 
$\log(P/{\rm days})=1.5$. 
In other words, the discrepancy between the pulsation
and the evolutionary mass decreases when moving from lower to higher mass
Cepheids and  disappears for masses around $13-14 M_{\odot}$. The authors
interpreted this discrepancy as an effect of mass loss occurring during
and before the central He-burning phase. However, the data would imply
a mass-loss efficiency that decreases with increasing Cepheid initial
mass; this mass-loss is at odds with empirical estimates 
where the mass loss increases with stellar luminosity 
and radius (Schroeder \& Cuntz 2007). 
Another plausible solution of the Cepheid mass discrepancy is given
by extra convective mixing; in this case, the relation ML is modified
because the luminosity of the He-burning phase is higher than in the 
canonical case and increases with an increasing amount of extra mixing.   
Chiosi et al. (1992) discussed the mass discrepancy of Cepheids by comparing
models incorporating either semiconvection or overshoot and determining
the pulsation mass from Cepheid models (Chiosi, Wood and Capitanio 1993).
In the case of models with overshoot, the discrepancy no longer exists for the
considered Cepheids in the young rich LMC cluster NGC 2157.
 
On the one side, mass loss, in an ad-hoc manner at least, offers a mechanism 
capable of
modifying the M-L relation by directly reducing the mass of a Cepheid.
On the other side,  according to a discussion by
Neilson and Lester (2008, 2009),
 considering winds driven not only by radiation, but including momentum input
from pulsation and shocks generated in the atmosphere, the calculated
mass-loss rate for Cepheids is enhanced and helps us to illuminate the issue of
infrared excess and the mass discrepancy,

After reanalysing the results of Caputo et al (2005), Keller (2008) rejects 
the mass loss hypothesis and reaches the conclusion that an increased internal 
mixing remains the most likely cause of the Cepheids mass discrepancy.

As an exercise, we considered the same sample of Cepheids used
in Fig. 14 (Cepheids in open clusters of Table 3 of Tammann et al
 2003). For a solar chemical composition ($Z=0.017$, $Y=0.27$) we computed
a dense grid of isochrones ($\Delta\log(t/{\rm yr})=0.05$), then
for each point (Cepheid) of the plane $M_V-(M_V-M_I)$, we obtained the 
minimum distance from the points which describe the totality of the
isochrones. This minimum singles out  the age and the evolutionary mass 
$M_{\rm e}$ of 
the Cepheid under consideration (with an uncertainty  related to the 
ranging of isochrones). The corresponding pulsation mass $M_{\rm p}$
was computed using the mass-dependent PLC relation for fundamental
pulsators (employing the observed quantities $M_I$ and $(M_V-M_I)$ of Table 2 
in  Caputo et al (2005)). The estimated pulsation and evolutionary masses are
presented in Fig. 15. In the same figure, the open symbols 
represent the pulsation masses computed with the PLC relation by Chiosi et al 
(1993) (we point out that the opacity tables used in Chiosi et al.
differ from those of Caputo et al. and those of the present paper).

\begin{figure}
 \resizebox{\hsize}{!}{\includegraphics{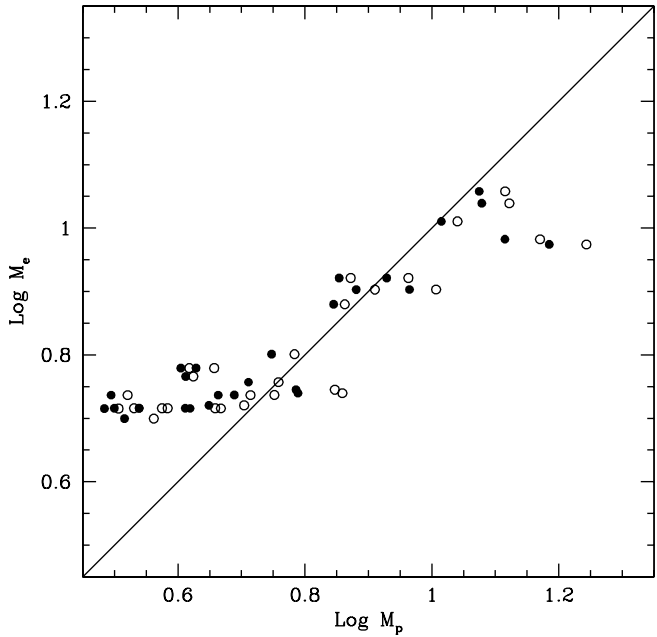}}
\caption{  
 Cepheids in open clusters and 
associations  (from the paper by Tammann et al. 2003) are plotted with the
pulsation and the evolutionary mass determined as described in the text. 
Solid dots are pulsation masses determined according to the PLC relation
by Caputo et al. (2005) and open circles according to the PLC relation by
Chiosi et al. (1993).  
 }
\label{massdis}
\end{figure}

Independently from the formulation used to obtain $M_{\rm p}$, the data plotted
 in this figure show that for stars of mass lower than $\sim 6 M_{\odot}$, the
evolutionary mass is higher than the pulsation one ($M_{\rm e} > M_{\rm p}$),
while for stars more massive than $10 M_{\odot}$ the pulsation mass is higher 
than the evolutionary one ($M_{\rm e} < M_{\rm p}$). 
In the range $6-10 M_{\odot}$, the parametrization $\Lambda_c=0.5$ could be 
satisfactory. We could interpret this  by suggesting that the parameter 
$\Lambda_c=0.5$, which determines the efficiency of the extramixing, must be a 
function of the mass of the star. In other words, $\Lambda_c$ should 
decrease with increasing stellar mass, assuming values close to $\Lambda_c=0.5$
 in the 
range $6-10 M_{\odot}$. This should be taken as a suggestion and not a firm 
result because the number of stars in the sample is very limited; the
values of the absolute magnitudes of the Cepheids are significantly different 
depending on the different estimates. 
 
From our sample, we excluded GY Sge and S Vul because their $M_{\rm p}$ 
was $> 20 M_{\odot}$ (i.e., beyond the validity range of theoretical 
relations), and EV Sct because it is probably a binary.
 The differences between 
the results of Keller and ours are mainly caused by
Keller proposing an additional internal mixing, parametrized by 
$\Lambda_c=0.67 \pm 0.17$ for the whole mass range, and us suggesting an extra 
mixing that decreases with the stellar mass, but with $\Lambda_c \approx 0.5$
in the range $6-10 M_{\odot}$.

\section{Massive red supergiants}

Red supergiants (RSGs) are a He-burning phase in the evolution of moderately
high mass stars ($10-25 M_{\odot}$). The evolution of these stars, particularly
at low metallicities, is still poorly understood. The latest-type RSGs in the
Magellanic Clouds are cooler than the current evolutionary tracks allow,
occupying the region to the right of the Hayashi limit where stars are no 
longer in hydrostaic equilibrium.  
Until recently, the location of Galactic red supergiants (RSGs) in the 
Hertzsprung-Russell (H-R) diagram was poorly matched by stellar evolutionary 
tracks,  evolutionary theory being unable to produce stars as cool and luminous
 as those observed  (Massey 2003). Many possible explanations might 
contribute to this discrepancy: there is poor knowledge of RSG molecular 
opacities, the near-sonic velocities of the convective layers invalidate 
simplifications of mixing length theory, and the highly extended atmospheres
of these stars differ from the plane-parallel geometry assumption adopted by
evolutionary models (Levesque et al. 2006). A similar problem was evident for 
RSGs in the Magellanic Clouds (MCs) as shown by the data of Massey \& Olsen 
(2003) with the more robust available calibration at that time. 

Afterwards Levesque et al. (2005, 2006) used moderate-resolution
optical spectrophotometry and the new MARCS stellar atmosphere models (Plez 
2003, Gustafsson et al. 2003) to determine the physical properties of RSGs in
both the Milky Way and the MCs for the comparison with the evolutionary models.
They derived a new effective temperature scale significantly warmer than those 
in the literature, showing that the newly derived temperatures and bolometric 
corrections provide closer agreement with stellar evolutionary tracks.
The comparison of Galactic RSGs was with the Geneva evolutionary tracks of 
solar metallicity (Meynet \& Maeder 2003) spanning a range of initial 
rotational velocities from $0$ to $300 km s^{-1}$. 
Levesque et al. (2006) extended their study to Magellanic Cloud red
supergiants to test the effects of metallicity and found evidence of 
significant visual extinction because of circumstellar dust. The effective 
temperatures of K supergiants are about the same in the SMC, LMC and 
Milky Way but the lower abundance of TiO leads to effective temperatures 
that are lower for M supergiants of the same spectral subtype.
On average, RSGs in the Magellanic Clouds are not as cool as Galactic RSGs, in
agreement with the shifting of the rightmost extension (the Hayashi limit)
of the evolutionary tracks to warmer effective temperatures at lower 
metallicities. The observed average spectral subtypes of RSGs in these galaxies
shifts from $M2 I$ in the Milky Way to $M1 I$ in the LMC and $K5-K7 I$ in the
SMC.
The newly derived physical parameters by Levesque et al. (2005, 2006) brought 
the location of the RSGs  into
much closer agreement with the predictions of stellar evolutionary theory.
There is now excellent agreement between Milky Way supergiants and the
evolutionary tracks, accurately reproducing this stage of massive star
evolution in the Milky Way. Figures 16, 17, and 18 show our more
massive tracks from 10 to 20 $M_{\odot}$ displaying also the data from Levesque
 et al. (2005, 2006) of  RSGs in the Milky Way, LMC, and SMC.

\begin{figure}
 \resizebox{\hsize}{!}{\includegraphics{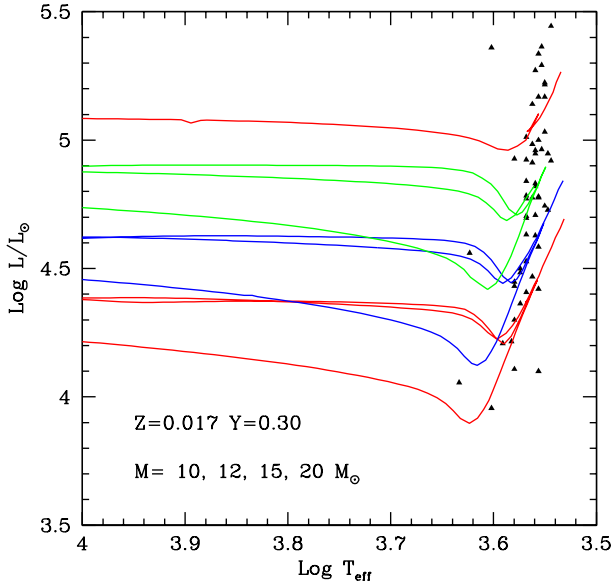}}
\caption{  
Evolutionary tracks in the HR diagram, for the composition $Z=0.017,
Y=0.30$ for masses $10, 12, 15$, and $20 M_{\odot}$. Triangles are the 
Galactic red supergiants from Levesque et al. (2005).
 }
\label{rsg_mw}
\end{figure}

\begin{figure}
 \resizebox{\hsize}{!}{\includegraphics{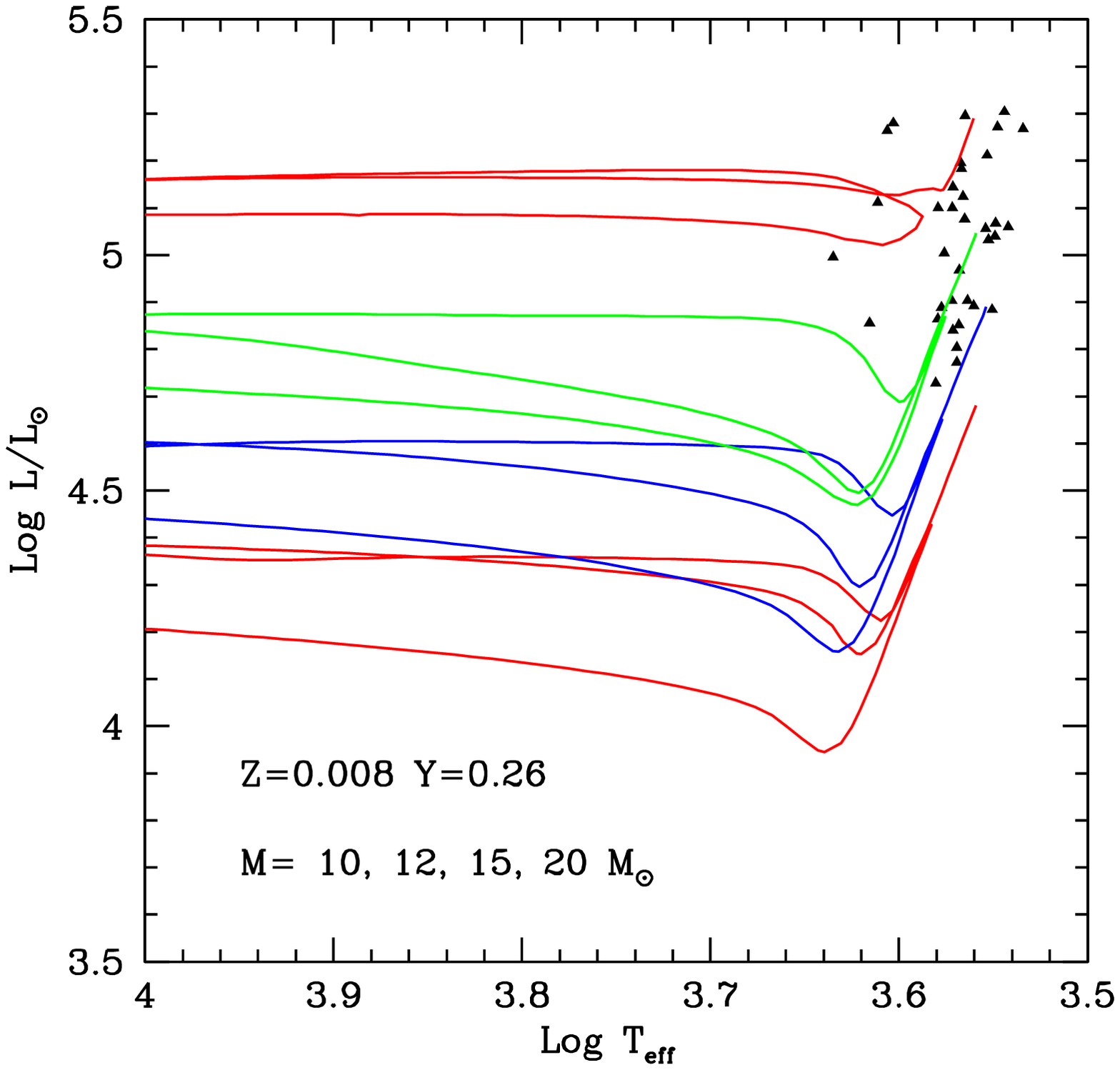}}
\caption{  
Evolutionary tracks in the HR diagram, for the composition $Z=0.008,
Y=0.26$ for masses $10, 12, 15$, and $20 M_{\odot}$. Triangles are the 
LMC red supergiants from Levesque et al. (2006).
 }
\label{rsg_lmc}
\end{figure}

\begin{figure}
 \resizebox{\hsize}{!}{\includegraphics{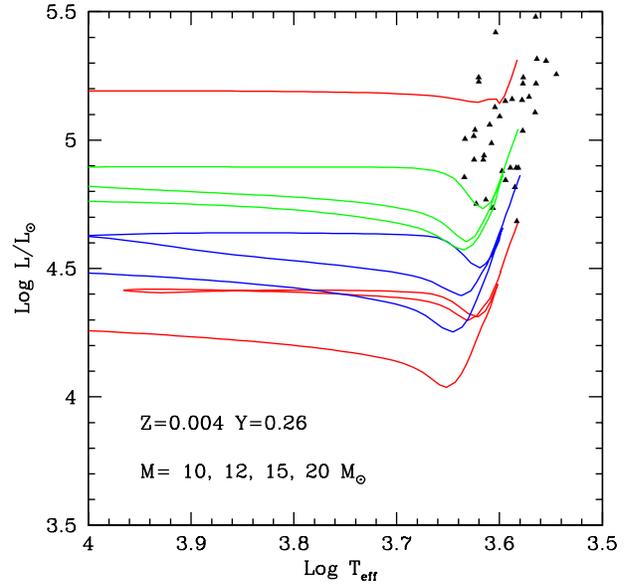}}
\caption{  
Evolutionary tracks in the HR diagram, for the composition $Z=0.004,
Y=0.26$ for masses $10, 12, 15$, and $20 M_{\odot}$. Triangles are the 
SMC red supergiants from Levesque et al. (2006).
 }
\label{rsg_smc}
\end{figure}

\section{Concluding remarks}
\label{sec_remarks}


Observations of some globular clusters in the Galaxy and the LMC provide 
evidence of multiple populations and/or  significant variations in the
helium content, in contrast to traditional scenarios in which
 globular clusters are
simple stellar populations of uniform age and chemical composition.
The current database of Padova tracks allows the computation of isochrone sets 
covering a wide range of ages and enables users to analyse stellar populations 
(star clusters or galaxies) with different helium enrichment laws, given
the extended $Z-Y$ region of the evolutionary models. 

The scaled solar database of models and isochrones, presented in Paper I, has 
been extended in mass to include stellar evolutionary computations from
$2.5$ to $20 M_{\odot}$ for the same large region of the $Z-Y$ plane.  
An important update of this database is the extension of stellar models and 
isochrones until the end of the TP-AGB phase by means of synthetic models,
using the same relations as in Marigo \& Girardi (2007).
Revised bolometric corrections by Girardi et al. (2008) are employed to 
transform Padova isochrones into UBVRIJHK and ACS photometric systems.

In the range of more massive stars presented in this Paper II, our models can be
tested in at least two interesting astrophysical problems: 1) Cepheids and the 
mass-discrepancy problem, 2) massive red supergiants and the problems of 
their effective temperature (Massey 2003).

Considering the observations of Cepheids in the Galaxy and the Magellanic 
Clouds 
(Tamman et al. 2003; Sandage et al. 2004; Sandage et al. 2009), we show that 
 the observed instability strip is narrower than the theoretical one, and that
the slope of the lower boundary of observed stars coincides with the slope of 
the faint envelope connecting each of the loops of the theoretical tracks for 
the chemical 
composition of the MCs. Because of the small number of Cepheids in open 
clusters and associations of the Milky Way, we cannot draw the lower boundary 
of the observed strip as we did for the MCs, but in this case, the loops 
of the theoretical models can also indicate the location of MW Cepheids in the 
HR-diagram.

The long-standing open question about the discrepancy between the pulsation and
the evolutionary mass of Cepheid stars is considered, by determining the 
evolutionary mass of Cepheids in MW open clusters (Tamman et al. 2003) from
our models (isochrones) and the pulsation mass from the mass dependent PLC
relation for fundamental pulsators. The evolutionary mass is found to be higher
than the pulsation mass for stars of mass lower than $\sim 6 M_{\odot}$, while 
for stars more massive than $10M_{\odot}$ the pulsation mass is higher than the
evolutionary one. If we consider extra convective mixing as a plausible 
solution of the Cepheid mass discrepancy, the parameter that determines the
efficiency of the extra mixing should decrease with increasing stellar mass.

Red supergiants in both the Galaxy and the Magellanic Clouds were too cool and
luminous compared with evolutionary tracks (Massey 2003; Massey \& 
Olsen 2003) until when Levesque et al. (2005, 2006) derived a new effective 
temperature scale and new bolometric corrections, producing far closer
agreement with stellar evolutionary tracks. Our tracks from $10$ to $20 
M_{\odot}$ for the appropriate chemical composition can reproduce the
location in the H-R diagram of red supergiants in the MW and the MCs as given 
by Levesque et al. (2005, 2006).

A web site has been dedicated to make available the entire database 
 to the scientific community.

 In the section {\bf YZVAR} of the {\bf static databases} 
at {\bf http://stev.oapd.inaf.it} users can find:

\begin{itemize}
\item
{data files with the information relative to each evolutionary track},
\item
{isochrone files from Paper I 
with the chemical composition of the computed grids in 
the Z-Y plane, and age interval in the range $\sim 10.15 \ge$  log age/yr 
$ \ge \sim 9.00$. }
\end{itemize}
In the section {\bf YZVAR form} of the {\bf interactive services} a web 
interface allows users to obtain interpolated isochrones for the entire
range of ages, from old (log age/yr $\sim 10.15$) to young ages (log age/yr 
$ \sim 7.00$) and any chemical composition in the provided range
of Y and Z.   
 The initial and final values of isochrone ages can vary a little 
more  or less, depending on the chemical composition.
To compute the isochrones we adopt the interpolation scheme described in 
Sect. 5.1.

We point out that older isochrones (Paper I) are available in both the static 
database 
and the interactive service, while younger isochrones only interactively.  
All isochrones (old and young) are available through the interactive service
for the chosen photometric system (see  Sect. 5.3).

In the future, a web interactive interface will provide data of stellar 
populations in terms of 
 SFR, IMF, chemical composition, and mass loss during the RGB phase.

\begin{acknowledgements}
We acknowledge our referee's useful comments and suggestions that helped to
improve the presentation of our results.
We thank  C. Chiosi for his continuous interest and support to stellar
evolution computations. We thank A. Weiss and B. Salasnich for help with 
opacity tables, G. Bono and A. Bressan for useful discussions.
We acknowledge financial support from INAF COFIN 2005 ``A Theoretical lab
for stellar population studies'' and  from
Padova University (Progetto di Ricerca di Ateneo CPDA 052212).

\end{acknowledgements}



\begin{thebibliography}{}
\bibitem[]{} Alexander, D.R. \& Ferguson, J.W. 1994, ApJ, 437, 879
\bibitem[]{} Allard, F. et al. 2000, in From giant planets to cool stars,
ASP Conf. Ser. 212, eds. C.A. Griffith \& M.S. Marley, p. 127 
\bibitem[]{} Alongi, M., Bertelli, G., Bressan, A., Chiosi, C. 1991, A\&A,
    244, 95 
\bibitem[]{} Alongi, M. et al. 1993 A\&AS, 97, 851
\bibitem[]{} Anders, E. \& Grevesse, N. 1989, Geochim.\ Cosmochim.\ Acta, 53, 19
\bibitem[]{} Angulo, C., Arnould, M., Rayet, M. et al. 1999, Nucl. Phys. A, 656, 3
\bibitem[]{} Asplund, M., Grevesse, N., Sauval, A.J. et al. 2004, A\&A, 417, 751
\bibitem[]{} Basu, S. \& Antia, H.M. 2008, PhR, 457, 217
\bibitem[]{} Bemmerer, D. et al. (LUNA Collaboration) 2006, Nucl. Phys. A, 779, 297 
\bibitem[]{} Bertelli, G., Bressan, A., Chiosi, C., Fagotto, F., Nasi, E. 
      1994, A\&AS, 106, 275
\bibitem[]{} Bertelli, G., Girardi, L., Marigo, P., Nasi, E. 2008, A\&A, 484,
815 (Paper I)
\bibitem[]{} Bessell, M.S. 1990, PASP, 102, 1181
\bibitem[]{} Bessell, M.S. \& Brett, J.M. 1988, PASP, 100, 1134
\bibitem[]{} B\"{o}hm-Vitense, E. 1958, ZA, 46, 108
\bibitem[]{} Bono, G., Caputo, F., Cassisi, S. et al. 2000, ApJ, 543, 955   
\bibitem[]{} Bressan, A., Bertelli, G., Chiosi, C. 1981, A\&A, 102, 25 
\bibitem[]{} Bressan, A., Fagotto, F., Bertelli, G., Chiosi, C. 1993, 
	A\&AS, 100, 647
\bibitem[]{} Caffau, E., Ludwig, H.-G., Steffen, M. et al. 2008, A\&A, 488, 1031
\bibitem[]{} Caputo, F. et al. 2004, A\&A, 424, 927
\bibitem[]{} Caputo, F. et al. 2005, A\&A, 629, 1021
\bibitem[]{} Castellani, V.  et al. 1985, ApJ, 296, 204
\bibitem[]{} Castellani, V., Chieffi, A. \& Straniero, O. 1990, ApJS 74, 463
\bibitem[]{} Castelli, F. \& Kurucz, R.L. 2003, in Modelling of 
Stellar Atmospheres, IAU Symp. 210, eds. N.E. Piskunov et al., p. 20 
(San Francisco: ASP)
\bibitem[]{} Castor, J.I., Abbott, D.C., Klein, R.I. 1975, ApJ, 195, 157 
\bibitem[]{} Caughlan, G.R. \& Fowler, W.A. 1988, Atomic Data Nucl.\ Data 
      Tables, 40, 283
\bibitem[]{} Charbonnel, C., Meynet, G., Maeder, A. et al. 1993, A\&A, 101, 415

\bibitem[]{} Chiosi, C., Bertelli, G., Bressan, A. 1992, ARA\&A, 30, 305
\bibitem[]{} Chiosi, C. \& Maeder, A. 1986, ARA\&A, 24, 239
\bibitem[]{} Chiosi, C., Wood, P.R., Bertelli, G., Bressan, A., Mateo, M. 
1992, ApJ, 385, 205
\bibitem[]{} Chiosi, C., Wood, P.R., Capitanio, N. 1993, ApJS, 86, 541
\bibitem[]{} Cordier, D., Goupil, M.J., Lebreton, Y. 2003, A\&A, 409, 491
\bibitem[]{} de Jager, C., Nieuwenhuijzen, H., van der Hucht, K. 1988, A\&AS,
72, 295
\bibitem[]{} de Mink, S.E., Cantiello, M., Langer, N. et al. 2009, A\&A, 497, 243 
489, 685
\bibitem[]{} Fagotto, F., Bressan, A., Bertelli, G., Chiosi, C. 1994a, A\&AS, 104, 365
\bibitem[]{} Fagotto, F., Bressan, A., Bertelli, G., Chiosi, C. 1994b, A\&AS, 105, 29
\bibitem[]{} Ferguson, J. W. et al. 2005, ApJ, 623, 585
\bibitem[]{} Formicola, A. et al.(LUNA Collaboration) 2004, PhLB, 591, 61
\bibitem[]{} Fluks, M.A. et al. 1994, A\&AS, 105, 311
\bibitem[]{} Girardi, L., Bressan, A., Chiosi, C., Bertelli, G., Nasi, E.
1996, A\&AS, 117, 113
\bibitem[]{}
Girardi, L., Bressan, A., Bertelli, G., Chiosi, C. 2000, A\&AS, 141, 371

\bibitem[]{} Girardi, L. et al. 2002, A\&A, 391, 195
\bibitem[]{} Girardi, L. et al. 2008, PASP, 120, 583
\bibitem[]{} Girardi, L., Castelli, F., Bertelli, G., Nasi, E. 2007, 
A\&A, 468, 657
\bibitem[]{} Graboske, H.C., de Witt, H.E., Grossman, A.S., Cooper, M.S. 1973,
        ApJ, 181, 457
\bibitem[]{} Grevesse, N., Asplund, M., Sauval, A.J. 2007, Space SciRev, 130, 105
\bibitem[]{} Grevesse, N. \& Sauval, A.J. 1998, Space Sci. Rev., 85, 161
\bibitem[]{} Grevesse, N. \& Noels, A. 1993, Phys. Scr. T, 47, 133
\bibitem[Grevesse(1991)]{1991JPhy4...1..181G} Grevesse, N.\ 1991, Journal 
de Physique IV, 1, 181 
\bibitem[]{} Gustafsson, B. et al. 2003, in ASP Conf. Ser. 288, Stellar 
Atmosphere Modeling, eds. I. Hubeny et al. , p. 331
\bibitem[]{} Guzik, J.A. et al. 2005, ApJ, 627, 1049
\bibitem[]{} Guzik, J.A. et al. 2006, MSAIt, 77, 389 
\bibitem[]{} Haft, M., Raffelt, G. \& Weiss, A. 1994, ApJ, 425, 222
\bibitem[]{} Heger, A. \& Langer, N. 2000, ApJ, 544, 1016 
\bibitem[]{} Holtzman, J.A. et al. 1995, PASP, 107, 1065
\bibitem[]{} Iglesias, C.A. \& Rogers, F.J.  1996, ApJ, 464, 943
\bibitem[]{} Imbriani, G. et al. 2004, A\&A, 420, 625
\bibitem[]{} Imbriani, G. et al. (LUNA Collaboration), 2005, EPJA, 25, 455
\bibitem[]{} Itoh, N.,Mitake, S., Iyetomi, H. \& Ichimara, S. 1983, ApJ, 273, 774
\bibitem[]{} Keller, S. C. 2008, ApJ, 677, 483
\bibitem[]{} Kippenhahn, R., Weigert, A., Hofmeister, E. 1967, in Methods in 
        Computational Physics, eds. B.\ Alder, S. Fernbach, M. Rotenberg, 
        New York: Academic Press, Vol.\ 7, p.\ 129
\bibitem[]{} Kudritzki, R.P. 2002, ApJ, 577, 389
\bibitem[]{} Kudritzki, R.P. et al. 1989, IAU Coll. 113, pag. 67
\bibitem[]{} Kudritzki, R.P. \& Puls,J. 2000, ARA\&A, 38, 613  
\bibitem[]{} Landr\'e, V., Prantzos, N., Aguer, P., Bogaert, G., Lefebvre, A., 
	Thibaud, J.P. 1990, A\&A, 240, 85
\bibitem[]{} Lemut, A. et al. (LUNA Collaboration) 2006, Phys. Lett, B, 634, 483
\bibitem[]{} Levesque, E. M., Massey, P., Olsen, K.A.G. et al. 2005,
 ApJ, 628, 973
\bibitem[]{} Levesque, E. M., Massey, P., Olsen, K.A.G. et al. 2006. 
 ApJ, 645, 1102
\bibitem[]{} Loidl, R., Lancon, A., Jorgensen, U.G. 2001, A\&A, 371, 1065

\bibitem[]{} Lucy, L.B. \& Solomon, P.M. 1970, ApJ, 159, 879
\bibitem[]{} Maeder, A. 2009, Physics, Formation and Evolution of Rotating
Stars, Springer Berlin Heidelberg, 2009

\bibitem[]{} Marigo, P. \& Aringer, B. 2009, arXiv:0907.3248
\bibitem[]{} Marigo, P. et al. 2001, A\&A, 371, 152
\bibitem[]{} Marigo, P.\& Girardi, L. 2007, A\&A, 469, 239
\bibitem[]{} Marigo, P. et al. 2008, A\&A, 482, 883
\bibitem[]{} Massey, P. 2003, ARA\&A, 41, 15 
\bibitem[]{} Massey, P. \& Olsen, K.A.G. 2003, AJ, 126, 2867
\bibitem[]{} Massey, P. et al. 2008, in Proc. of IAU Symp. 250, p. 97
\bibitem[]{} Meynet, G. \& Maeder, A. 2000, A\&A, 361, 101
\bibitem[]{} Meynet, G., Maeder, A., Schaller, G., Schaerer, D., Charbonnel,
 C. 1994, A\&AS, 103, 97
\bibitem[]{} Mihalas, D., Hummer, D.G., Mihalas, B.W., D\"appen, W. 1990, ApJ,
	350, 300
\bibitem[]{} Mokiem, M.R.  et al. 2007, A\&A, 473, 603
\bibitem[]{} Montalban, J. et al. 2004, ESASP, 559, 574
\bibitem[]{} Neilson, H.R., Lester, J.B. 2008, ApJ, 684, 569
\bibitem[]{} Neilson, H.R. et al. 2009, ApJ, 690, 1829
\bibitem[]{} Pomp\'eia, L., Hill, V., Spite, M. et al. 2008, A\&A, 480, 379
\bibitem[]{} Pietrinferni, A., Cassisi, S., Salaris, M.,  Castelli, F. 2004,
  ApJ, 612, 168
\bibitem[]{} Pietrinferni, A., Cassisi, S., Salaris, M.,  Castelli, F. 2006,
  ApJ, 642, 797
\bibitem[]{} Piotto, G. et al. 2005, ApJ, 621, 777
\bibitem[]{} Piotto, G., Bedin, L. R., Anderson, J. et al. 2007, ApJ, 661, L53
\bibitem[]{} Piotto, G. 2009, in IAU Symp. 258, The Ages of Stars, p.233
\bibitem[]{} Plez, B. 2003, in ASP Conf. Ser. 298, GAIA Spectroscopy: Science
and Technology, ed. U. Munari,  p. 189
\bibitem[]{} Rogers, F.J., Swenson, F.J., Iglesias, C.A. 1996, ApJ, 456, 902
\bibitem[]{} Romaniello, M. et al. 2008, A\&A, 488, 731
\bibitem[]{} Salaris, M., Chieffi, A., Straniero, O. 1993, ApJ, 414, 580
\bibitem[]{} Salasnich, B., Girardi, L., Weiss, A.,  Chiosi, C. 2000, 
A\&A, 361, 1023
\bibitem[]{} Salpeter, E.E. 1955, ApJ, 121,161
\bibitem[]{} Sandage, A., Tammann, G.A. \& Reindl, B. 2004, A\&A, 424, 43
\bibitem[]{} Sandage, A., Tammann, G.A. \& Reindl, B. 2009, A\&A, 493, 471
\bibitem[]{} Schaerer, D. et al. 1993a, A\&AS, 98, 523
\bibitem[]{} Schaerer, D. et al. 1993b, A\&AS, 102, 339
\bibitem[]{} Schaller, G. et al. 1992, A\&AS, 96, 269
\bibitem[]{} Schlattl, H. \& Weiss, A. 1998, in Proc. Neutrino Astrophysics,
Ringberg Castle, Germany, Oct 1997, eds. M. Altmann, W. Hillebrandt et al.   
\bibitem[]{} Schroder, K.P. \& Cuntz, M. 2007, A\&A, 465, 593  
\bibitem[]{} Scott, P., Asplund, M., Grevesse, N., Sauval, A.J. 2009,
ApJ, 691, L119
\bibitem[]{} Sirianni, M. et al. 2005, PASP, 117, 1049
\bibitem[]{} Storm, J. et al. 2004, A\&A, 415, 531
\bibitem[]{} Tammann, G.A., Sandage, A. \& Reindl, B. 2003, A\&A, 404, 423
\bibitem[]{} Tolstoy, E., Hill, V., Tosi, M. 2009, ARA\&A in press, astro-ph/0904.4505
\bibitem[]{} Vemury, S.K., Stothers, R. 1978, ApJ, 225, 939
\bibitem[]{} Villanova,S., Piotto, G., King, I. et al. 2007, ApJ, 663, 296
\bibitem[]{} Vink, J.S., de Koter, A., Lamers, H.J.G.L.M. 2001, A\&A, 369, 574
\bibitem[]{} Vink, J.S. \& de Koter, A. 2005, A\&A, 442, 587
\bibitem[]{} Weaver T.A. \& Woosley S.E. 1993, Phys. Rep., 227, 65
\bibitem[]{} Weiss, A., Keady, J.J., Magee, N.H. Jr. 1990, Atomic Data and 
Nuclear Data Tables, 45, 209
\bibitem[]{} Weiss, A. \& Schlattl, H. 2000, A\&AS, 144, 487
\bibitem[]{} Weiss, A., Serenelli,A., Kitsikis, A. et al. 2005, A\&A, 441, 1129
\bibitem[]{} Yi, S., Demarque, P., Kim, Y.-C. et al. 2001, ApJS, 136, 417 


\end{thebibliography}
\end{document}